\documentclass[aps, prd, reprint, 
10pt, notitlepage, a4paper,
floats, floatfix,
amsmath, amssymb, amsfonts,
superscriptaddress,
showpacs, showkeys,
nofootinbib,
longbibliography,
]{revtex4-2}

\usepackage{graphicx} % include figures
\usepackage{grffile}
\usepackage[usenames,dvipsnames]{xcolor}
\usepackage{xspace} % Sensible space treatment at end of simple macros
\usepackage{bm} % bold math
\usepackage{amsthm}
\usepackage[utf8]{inputenc} % for some references from inspirehep.net
\usepackage{booktabs}
\usepackage{enumerate}
\usepackage[normalem]{ulem} % for \sout

\xdefinecolor{mylinkcolor}{rgb}{0,0,0.5}
\usepackage[
	bookmarksnumbered, bookmarksopen, bookmarksopenlevel=2,
	breaklinks=true,
	colorlinks=true, filecolor=mylinkcolor, citecolor=mylinkcolor,
	linkcolor=mylinkcolor, urlcolor=mylinkcolor, menucolor=mylinkcolor,
]{hyperref}
\usepackage{orcidlink}

\graphicspath{ {./images/} }

\def\di{\partial}
\def\mF{\mathcal{F}}

\def\pphi{p_\varphi}
\def\pphidot{\dot{p}_\varphi}

\allowdisplaybreaks

\newcommand{\AEI}{\affiliation{Max Planck Institute for Gravitational Physics (Albert Einstein Institute), Am M\"uhlenberg 1, Potsdam 14476, Germany}}
\newcommand{\NBIA}{\affiliation{Niels Bohr International Academy, Niels Bohr Institute, Blegdamsvej 17, 2100 Copenhagen, Denmark}}
\newcommand{\URI}{\affiliation{Department of Physics and Center for Computational Research, University of Rhode Island, Kingston, RI 02881, USA}}  
\newcommand{\UMassDPhy}{\affiliation{Department of Physics, University of Massachusetts, Dartmouth, MA 02747, USA}}
\newcommand{\CSCVRUMass}{\affiliation{Center for Scientific Computing and Data Science Research, University of Massachusetts, Dartmouth, MA 02747, USA}}
\newcommand{\PI}{\affiliation{Perimeter Institute for Theoretical Physics, 31 Caroline Street North, Waterloo, ON N2L 2Y5, Canada}}
\newcommand{\UMD}{\affiliation{Department of Physics, University of Maryland, College Park, MD 20742, USA}}

\begin{document}

\title{Testing eccentric corrections to the radiation-reaction force in the test-mass limit of effective-one-body models}

\author{Guglielmo Faggioli}
\email{guglielmo.faggioli@aei.mpg.de}
\AEI

\author{Maarten van de Meent}
\NBIA
\AEI

\author{Alessandra Buonanno}
\AEI\UMD

\author{Aldo Gamboa}
\AEI

\author{Mohammed Khalil}
\PI

\author{Gaurav Khanna}
\URI
\UMassDPhy
\CSCVRUMass

% ----------------------------------------
% --------------- Abstract ---------------
% ----------------------------------------

\begin{abstract}
  In this work, we test an effective-one-body radiation-reaction force for eccentric 
  planar orbits of a test mass in a Kerr background, which contains third-order post-Newtonian (PN) non-spinning and second-order PN spin contributions. 
  We compare the analytical fluxes connected to two different resummations of this force, truncated at different PN orders in the eccentric sector, with the numerical fluxes 
  computed through the use of frequency- and time-domain Teukolsky-equation codes.
  We find that the different PN truncations of the radiation-reaction force show the expected scaling in the weak gravitational-field regime, and we observe a fractional difference with the numerical fluxes that is $<5 \%$, 
  for orbits characterized by eccentricity $0 \le e \le 0.7$, central--black-hole spin $-0.99 M \le a \le 0.99 M$ and fixed orbital-averaged quantity $x=\langle M\Omega \rangle^{2/3} = 0.06$, corresponding to the mildly strong-field regime  
  with semilatera recta $9 M<p<17 M$.
  Our analysis provides useful information for the development of spin-aligned eccentric models in the comparable-mass case.
\end{abstract}

\maketitle
% ----------------------------------------
% ------------- Introduction -------------
% ----------------------------------------
\section{Introduction} \label{Sec: Introduction}

The observation of gravitational waves (GWs) with the LIGO-Virgo~\cite{LIGOScientific:2018mvr, LIGOScientific:2020ibl, LIGOScientific:2021usb} and LIGO-Virgo-KAGRA (LVK)~\cite{LIGOScientific:2021djp} collaborations marked a new era in 
gravitational physics, uncovering unique properties of stellar-mass black holes (BHs) and neutron stars. 
As future data acquisition becomes characterized by increased sensitivity, it is
necessary to improve the precision and accuracy of waveform models used for matched-filtering and parameter-estimation pipelines. In particular, modeling waveforms 
from eccentric and precessing-spin binaries will become increasingly important. This is further motivated by the fact that upcoming observational runs~\cite{KAGRA:2013rdx} and future
detectors, like the Einstein Telescope~\cite{Punturo_2010}, Cosmic Explorer~\cite{Evans:2021gyd} and LISA~\cite{LISA}, will increase the number of detections by a factor 
$\sim 10^3$ and be able to probe binary's subpopulations, in lower frequency bands and for smaller mass ratios, which  
can exhibit larger eccentricities~\cite{Hild:2010id, LIGOScientific:2016wof, Babak:2017tow, Gair:2017ynp}. 

While eccentricity decreases toward merger due to the energy and angular momentum loss caused by the emission of GWs~\cite{Peters:1963ux, Peters:1964zz},
the residual eccentricity can help constrain 
different binary-formation scenarios and thus the origin of GW sources~\cite{Mandel:2009nx, Rodriguez:2018rmd, Fragione:2018vty, Zevin:2021rtf}. 
Indeed, the eccentricity is indicative of binaries formed through dynamical formation channels, which could occur in dense stellar environments, like globular clusters, where the three-body Kozai-Lidov mechanism~\cite{Kozai:1962zz, LIDOV1962719} or dynamic capture~\cite{Samsing:2013kua, Zevin:2018kzq, Zevin:2021rtf,PortegiesZwart:1999nm, Miller:2001ez} play a role. 
Efforts are currently underway to detect signs of orbital eccentricity in the GW signals 
observed by the LVK detectors~\cite{Romero-Shaw:2019itr, Romero-Shaw:2020thy, Romero-Shaw:2021ual, Romero-Shaw:2022xko, Gamba:2021gap, Gayathri:2020coq,  Bonino:2022hkj, Iglesias:2022xfc, Clarke:2022fma, Knee:2022hth,Ramos-Buades:2023yhy}. 

Among the different methods to solve the two-body problem in general relativity, the effective-one-body (EOB) approach~\cite{Buonanno:1998gg, Buonanno:2000ef} is a 
framework that provides accurate and fast waveforms for quasi-circular (QC) binaries~\cite{Damour:2008gu, Pan:2010hz, Pan:2013rra, Taracchini:2013rva, Bohe:2016gbl, Nagar:2018zoe, Cotesta:2018fcv, Babak:2016tgq, Ossokine:2020kjp, Nagar:2018gnk, Riemenschneider:2021ppj, Pompili:2023tna}, due to a strong synergy between analytical approximation methods and numerical relativity (NR) results.

In recent years, generalizations of EOB models to eccentric inspirals have been developed~\cite{Bini:2012ji, Hinderer:2017jcs, Cao:2017ndf, Liu:2019jpg, Liu:2021pkr, Liu:2023dgl, Chiaramello:2020ehz, Nagar:2021gss, Albanesi:2021rby, Placidi:2021rkh, Khalil:2021txt, Ramos-Buades:2021adz}. In particular, Ref.~\cite{Khalil:2021txt} derived the second-order post-Newtonian (PN) expressions for the radiation-reaction (RR) force and gravitational-waveform modes 
for eccentric inspirals. They include tail effects, in addition to spin-orbit (SO) and spin-spin (SS) couplings. 
Reference~\cite{Ramos-Buades:2021adz} introduced the \texttt{SEOBNRv4EHM} model: an extension of the QC \texttt{SEOBNRv4HM} model~\cite{Cotesta:2018fcv} to eccentric orbits, where the authors showed an EOB/NR unfaithfulness less than $1\%$ when comparing with the 28 eccentric NR simulations that were publicly available at the time from the Simulating eXtreme Spacetimes (SXS) collaboration~\cite{Boyle:2019kee,SXS:catalog}.
\footnote{
The \emph{unfaithfulness} is a metric that quantifies the disparity between two waveforms as observed by a GW detector (a lower unfaithfulness indicates greater similarity between the waveforms).
The definition of the unfaithfulness is presented in, e.g., Ref.~\cite{Ramos-Buades:2021adz}.
To determine the accuracy of the \texttt{SEOBNRv4EHM} model, Ref.~\cite{Ramos-Buades:2021adz} employed an optimization over eccentricity and starting frequency to find the best-fitting EOB waveform  corresponding to a given eccentric NR waveform.
Thus, the \texttt{SEOBNRv4EHM} model showed an unfaithfulness against NR waveforms less than $1\%$ for systems with eccentricities below $ \sim 0.3 $.
}
However, in the work of Ref.~\cite{Ramos-Buades:2021adz}, the dynamics of the binary is modeled through the use of the QC RR force from the \texttt{SEOBNRv4HM} model, and the eccentric corrections are considered only when computing the eccentric waveforms modes introduced in Ref.~\cite{Khalil:2021txt}. By contrast, the more recent waveform model \texttt{SEOBNRv5EHM} employs a RR force with eccentric corrections~\cite{Gamboa:2024imd,Gamboa:2024hli}.

Among the other examples of eccentric EOB models, we mention \texttt{TEOBResumS}~\cite{Nagar:2020pcj, Riemenschneider:2021ppj}, which has been extended to eccentric orbits, after investigating several prescriptions for incorporating eccentricity effects, in Refs.~\cite{Chiaramello:2020ehz, Nagar:2021gss, Placidi:2021rkh, Nagar:2021xnh, Albanesi:2021rby, Albanesi:2022ywx, Albanesi:2022xge, Nagar:2022fep, Albanesi:2023bgi, Placidi:2023ofj, Nagar:2024dzj}. 
The latest version of their model, known as \texttt{TEOBResumS-Dal\`i}, includes eccentric 2PN information, and is characterized by factorizing the leading PN order of the waveform modes and azimuthal component of the RR force, which include high-order time derivatives of the radial separation and orbital frequency.
The eccentric 2PN radial component of the RR force is adapted from Ref.~\cite{Bini:2012ji} and is Pad\'e resummed.
When compared against the 28 SXS publicly-available eccentric NR simulations, the \texttt{TEOBResumS-Dal\`i} model shows unfaithfulness around $\sim 0.1\%$, although there are some NR waveforms for which the unfaithfulness is close to, or above $\sim 1\%$ \cite{Nagar:2024oyk}.
However, note that these unfaithfulness values cannot be directly compared against the ones computed in Ref.~\cite{Ramos-Buades:2021adz} for the \texttt{SEOBNRv4EHM} model.
This is because Refs.~\cite{Ramos-Buades:2021adz} and \cite{Nagar:2024oyk} employ different prescriptions for determining the EOB waveform that corresponds to a given NR simulation.
A direct comparison of these models employing the same prescription is presented in Ref.~\cite{Gamboa:2024hli}.

Several studies~\cite{Nagar:2006xv, Damour:2007xr, Damour:2008gu, Barausse:2011kb, Taracchini:2013wfa, Taracchini:2014zpa, Albanesi:2021rby, Albanesi:2022ywx, Albanesi:2023bgi} 
showed the importance of augmenting EOB waveform models for the plunge-merger and ringdown with insights from BH-perturbation theory. 
The common approach is to consider a test mass (TM) orbiting or scattering off a Kerr BH and use this system as a laboratory to test and provide benchmarks to the models in the comparable-mass 
case.
Among these works, Refs.~\cite{Albanesi:2021rby,Albanesi:2022ywx} assessed different EOB eccentric RR force prescriptions (either with distinct RR gauge choices or different factorizations of the RR force). 
In particular, Ref.~\cite{Albanesi:2022ywx} also provides an analysis for a proxy to the \texttt{SEOBNRv4HM} QC RR force and of a resummed version of the 2PN eccentric RR force introduced in Ref.~\cite{Khalil:2021txt}. (We will discuss some comparisons with their results in Sec.~\ref{Subsec: Instantaneous fluxes}.)

Here, we aim to extend past analyses in different ways. 
We consider the TM limit of a 3PN-eccentric RR force, recently computed by some of the authors of this work in Ref.~\cite{Gamboa:2024imd}. This force is computed employing the same procedure of Ref.~\cite{Khalil:2021txt}, but it considers a different  
gauge choice for the leading-order of the RR force, which avoids a
2.5PN modification (relative to the leading-order) of the
QC orbital phase when transforming between harmonic
and EOB coordinates. 
The force that we consider contains the full non-spinning contributions up to 3PN order, and spin contributions only up to 2PN order.
For the spin contributions, we employ the 1.5PN spin-orbit (SO) and 2PN spin-spin (SS) parts to the RR force as in Ref.~\cite{Khalil:2021txt}, but taking into account the leading-order gauge choice of Ref.~\cite{Gamboa:2024imd}.
We resum the RR force in two ways: by extracting  the QC RR force of \texttt{SEOBNRv5HM}~\cite{Pompili:2023tna} as a multiplicative and an additive term. In the TM limit, we remark that this QC RR force differs from the \texttt{SEOBNRv4HM} one by new higher-order PN contributions in the waveform modes used for the flux computation, as explained in Ref.~\cite{Pompili:2023tna}. 
We study the effects of the individual PN eccentric contributions to the RR force in different gravitational-field regimes of the parameter space.
The analysis is performed by comparing the analytical fluxes, computed from the eccentric RR force in the TM limit, against numerical fluxes that are computed by solving the Teukolsky equation~\cite{Teukolsky:1973ha} through the use of 
a frequency-domain (FD)~\cite{vandeMeent:2015lxa} and a time-domain (TD)~\cite{Sundararajan:2007jg,Sundararajan:2008zm, Field:2020rjr} code. 
Both fluxes are computed on equatorial geodesics of the Kerr metric and extracted at future null infinity.
By comparing the fluxes, we test the two resummations of the eccentric RR force, as we explain in detail in Sec.~\ref{Sec. II}. 
In our study, we focus on equatorial bound orbits of Kerr in the weak and strong gravitational-field regimes, and also explore the RR force for hyperbolic encounters, by comparing the fluxes for some Schwarzschild hyperbolic geodesics with fixed energy. 

The article is structured as follows. In Sec.~\ref{Sec. II}, we introduce the methodology of our analysis. In particular, Sec.~\ref{Subsec:EOB model in the TM limit} describes the EOB model we use in the TM limit and how we compute eccentric-planar geodesics in the Kerr metric. 
In Sec.~\ref{Subsec: Numerical fluxes}, we describe how to compute the numerical fluxes by solving the Teukolsky equation numerically, while Sec.~\ref{Subsec: Analytical fluxes and Schott terms} shows how the analytical fluxes are computed from the RR force through the use of the balance equations. In Sec.~\ref{Sec: Results}, we provide the main results of our analysis. In particular, Sec.~\ref{Subsec: Instantaneous fluxes} gives an overview of the fluxes comparison, and Secs.~\ref{Subsec: Bound orbits in Schwarzschild},~\ref{spinning orbits} show the results for bound orbits in the Schwarzschild and Kerr spacetime,  respectively. 
Section~\ref{Unbound orbits} focuses on hyperbolic encounters in Schwarzschild spacetime. 
Finally, Sec.~\ref{Conclusions} summarizes the results, and points out future steps. 
In the appendices, we provide supplemental information. Notably, in Appendix~\ref{Appendix A} we summarize how the non-spinning 3PN terms 
of the eccentric RR force are derived and we provide the full TM expressions of the eccentric corrections to the QC RR force, 
together with the expressions of the Schott terms, which are necessary to compute the instantaneous fluxes.
Appendix~\ref{App:FDInstFlux} provides details on how the instantaneous fluxes are computed from the FD Teukolsky-equation code.

% ------------- Notation --------------
\subsection*{Notations}

We adopt natural units  $G = c = 1$ and consider a non-spinning TM of mass $\mu = \nu M$ orbiting a Kerr BH of mass $M$ with dimensionless spin $a = J/M^2$.
The Kerr metric is expressed in Boyer-Lindquist coordinates $\{T, R, \theta, \varphi \}$ and we restrict our analysis to the equatorial plane, $\theta = \pi/2$.
The dynamics of the TM is described by canonical coordinates $\{ R, \varphi, P_R, P_{\varphi} \}$.
Throughout this article, we consider scaled dimensionless variables
\begin{gather}
\label{Mscaling}
  t = \frac{T}{M}, \quad r = \frac{R}{M}, \quad p_r = \frac{P_R}{\mu}, \quad 
  p_{\varphi}=\frac{P_\varphi}{M\mu}.
\end{gather}
The Hamiltonian $H$, and RR force $\mF = (\mF_r, \mF_{\varphi})$
are scaled by the TM $\mu$.

% ----------------------------------------

% ------------- Methodology --------------
% ----------------------------------------
\section{Methodology} \label{Sec. II}

In this work, we assess the analytical EOB eccentric RR force of Ref.~\cite{Khalil:2021txt}, extended to 3PN in the non-spinning part in Ref.~\cite{Gamboa:2024imd}, by comparing it with numerical results. In particular,
as we explain in Sec.~\ref{Subsec: Analytical fluxes and Schott terms}, we compare numerical fluxes obtained by solving the Teukolsky equation against the analytical fluxes that are connected to the RR force through the energy and angular-momentum balance equations.
These fluxes are computed on Kerr equatorial geodesics of a TM.
In the following two sections, we describe how the orbits are computed and the methodology used to derive the numerical and analytical fluxes.

\subsection{EOB model in the TM limit} \label{Subsec:EOB model in the TM limit}
  To describe the dynamics of a TM orbiting a Kerr BH in the equatorial plane, we work within the EOB framework~\cite{Buonanno:1998gg,Buonanno:2000ef} and consider the Kerr Hamiltonian restricted to equatorial orbits ($\theta = \pi/2$, $p_{\theta} = 0$): 
  \begin{equation} \label{Eq: Kerr_Hamiltonian}
    H = \Lambda^{-1} \left(2 a p_{\varphi}+\sqrt{\Delta  p_{\varphi}^2 r^2 + \Delta^2 \Lambda  \frac{p_r^2}{r}+\Delta  \Lambda  r} \right) \ , 
  \end{equation}
  with quantities $\Lambda$ and $\Delta$ being 
    \begin{subequations}
    \begin{align}
    \Lambda & = r^3 + 2 a^2 + a^2 r \ , \\
    \Delta & = a^2 - 2 r + r^2 \ .
    \end{align}
  \end{subequations}
  Instead of the radial momentum $p_r$ we consider $p_{r_{*}}$, which is the momentum conjugate to the tortoise radial coordinate $r_{*}$. 
  The tortoise coordinate is related to the Boyer-Lindquist coordinate $r$ by:
  \begin{subequations}
    \begin{align}
    & dr_{*} = \frac{r^2 + a^2}{\Delta} dr = \frac{1}{\xi(r)} dr \ , \\
    & p_{r_{*}} = \xi(r) p_r \ .
    \end{align}
  \end{subequations}
  This is a general practice~\cite{Damour:2007xr,Pan:2009wj} that is done to improve the numerical stability of the dynamical evolution, since 
  $p_r$ diverges at the horizon while $p_{r_{*}}$ does not.
  The evolution of the dynamics is provided by the Hamilton equations:
  \begin{subequations} \label{Ham_EOM}
    \begin{align}
      & \dot{r} = \xi \frac{\di H}{\di p_{r_{*}}}(r, p_{r_{*}}, p_{\varphi}) \ , \label{Ham_EOM_1} \\
      & \dot{\varphi} =  \Omega = \frac{\di H}{\di \pphi} (r, p_{r_{*}}, p_{\varphi}) \ , \label{Ham_EOM_2} \\ 
      & \dot{p}_{r_*} = -  \xi \frac{\di H}{\di r}(r, p_{r_{*}}, p_{\varphi}) + \mF_r \ , \label{Ham_EOM_3} \\
      & \pphidot = \mF_{\varphi} \ , \label{Ham_EOM_4}
    \end{align}
  \end{subequations} 
  where the dot symbol represents a total derivative with respect to the scaled coordinate time in Eq.~\eqref{Mscaling},
  $\Omega$ is the orbital frequency, scaled by the total mass, and $\mF=(\mF_r, \mF_{\varphi})$ corresponds to the RR force connected to the emission of GWs for generic equatorial orbits. 

  The RR force components we consider, $\mF_r$ and $\mF_{\varphi}$, are two resummed versions of the RR force originally computed in Ref.~\cite{Khalil:2021txt} and here extended to 3PN order in the non-spinning eccentric sector~\cite{Gamboa:2024imd}. 
  These two resummations are given by:
  \begin{subequations} \label{ecc impl}
  \begin{align} 
  & \mF_r^{\text{mult}} = \mF_r^{\text{QC}}\mF_r^{\text{ecc,mult}}  \quad , \quad \mF_{\varphi}^{\text{mult}}=\mF_{\varphi}^{\text{QC}}\mF_{\varphi}^{\text{ecc,mult}} \label{multiplicative impl} \\ 
  & \mF_r^{\text{add}} = \mF_r^{\text{QC}} + \mF_r^{\text{ecc,add}}  \quad , \quad \mF_{\varphi}^{\text{add}}=\mF_{\varphi}^{\text{QC}} + \mF_{\varphi}^{\text{ecc,add}} \label{additive impl} \ ,
  \end{align}
  \end{subequations}
  where $\mF_r^{\text{QC}}$ and $\mF_{\varphi}^{\text{QC}}$ are the radial and azimuthal components of the RR force using the QC prescription, while $\mF_{r, \varphi}^{\text{ecc,mult}}$ and $\mF_{r, \varphi}^{\text{ecc,add}}$ are the eccentric corrections.
  We refer to the two resummations as the \textit{multiplicative} implementation, given in Eq.~(\ref{multiplicative impl}) and the \textit{additive} implementation, given in Eq.~(\ref{additive impl}).
  The complete expressions in the TM limit of the eccentric corrections are in Appendix~\ref{Appendix A}. Note that we restrict our attention to resummations that reduce to \texttt{SEOBNRv5HM} in the QC limit. This precludes considering resummations of the type introduced in~\cite{Chiaramello:2020ehz}, which also change the QC radiation reaction.

  The QC RR force $\mF^{\text{QC}}=(\mF_r^{\text{QC}}, \mF_{\varphi}^{\text{QC}})$ is calculated using the prescription and PN information of the \texttt{SEOBNRv5HM} waveform model~\cite{Pompili:2023tna}, defined by the expressions
  \begin{subequations} \label{QC force}
  \begin{align} 
    & \mF_{\varphi}^{\text{QC}} = - \frac{\Omega}{8 \pi} \sum_{\ell = 2}^{8} \sum_{m = 1}^{\ell} m^2 |d_{\rm L} h_{\ell m}^{\rm F}|^2  \ , \label{azimuthal QC force} \\
    & \mF_{r}^{\text{QC}} = \frac{p_{r_{*}}}{p_{\varphi}} \mF_{\varphi}^{\text{QC}} \ \label{radial QC force},
  \end{align}
  \end{subequations}
  where $d_{\rm L}$ is the luminosity distance of the binary to the observer and $h_{\ell m}^{\rm F}$ are the GW modes in factorized form~\cite{Damour:2007yf, Damour:2008gu, Damour:2007xr, Pan:2010hz}, given by:
  \begin{equation}
    h_{\ell m}^{\rm F} = h_{\ell m}^{\rm (N, \epsilon)} \hat{S}_{\rm eff}^{(\epsilon)} T_{\ell m} f_{\ell m} e^{i \delta_{\ell m}} \label{fact_modes}\ .
  \end{equation}
  Here, $\epsilon$ is the parity of the multipolar waveform mode, such that $\epsilon = 0$ for even $\ell+m$, and $\epsilon = 1$ for odd $\ell+m$. The leading term in Eq. \eqref{fact_modes}, $h_{\ell m}^{\rm (N, \epsilon)}$ is the Newtonian contribution
  \begin{equation} \label{Newtprefactor}
    h_{\ell m}^{\rm (N, \epsilon)} = \frac{\nu}{d_L}n_{\ell m}^{(\epsilon)}c_{\ell+\epsilon}(\nu) v_\Omega^{\ell} Y^{\ell - \epsilon, -m}\left(\frac{\pi}{2}, \phi \right) \ ,
  \end{equation}
  where $Y^{\ell - \epsilon, -m}(\theta, \phi)$ are the scalar spherical harmonics, $n_{\ell m}^{(\epsilon)}$ and $c_{\ell+\epsilon}(\nu)$ are functions given in Eqs. (28) and (29) of Ref.~\cite{Pompili:2023tna}, and $v_\Omega$ is given by
  \begin{equation}
    v_{\Omega}=\Omega^{1/3}.
  \end{equation} 
Note that the \texttt{SEOBNRv5HM} model employs the variable 
\begin{equation}
\label{eq:vphi}
v_\varphi \equiv \Omega \left. \left(\frac{\partial H_{\rm EOB}}{\partial p_\phi} \right)^{-2/3}\right |_{p_r=0} \hspace{-0.5cm},
\end{equation}
instead of $v_\Omega$ in Eq.~\eqref{Newtprefactor}, which simplifies to $v_{\varphi} =  \Omega \, [ r^{3/2} + a ]^{2/3}$ in the TM limit.
However, the generalization of Eq.~\eqref{eq:vphi} to eccentric orbits is not straightforward due to the $ p_r = 0 $ requirement, which would affect the PN expansion of the eccentric modes unless a correction factor is applied.
Thus, by choosing $v_\Omega$ instead of  $v_{\varphi}$, one obtains a prescription that is easily applicable in generic orbits.
Furthermore, this choice does not have a significant impact on the accuracy of the underlying QC model \texttt{SEOBNRv5HM} \cite{Gamboa:2024hli}.
Therefore, the eccentric 3PN RR force in Eqs.~\eqref{Ham_EOM_3} and \eqref{Ham_EOM_4} is computed by employing $v_\Omega$.
  
  The function $\hat{S}_{\rm eff}^{(\epsilon)}$ is the effective source term which is given by 
  \begin{equation}
    \hat{S}_{\rm eff}^{(\epsilon)} = \begin{cases}
      H(r, p_{r_{*}}, p_{\varphi}), \ \ \epsilon = 0 \\
      p_{\varphi} v_\Omega, \ \ \ \ \epsilon = 1 \ .
  \end{cases}
  \end{equation}
  The factor $T_{\ell m}$ resums the leading order logarithms of tail effects and corresponds to
  \begin{equation}
    T_{\ell m} = \frac{\Gamma(\ell + 1 -2i\hat{k})}{\Gamma(\ell + 1)}e^{\pi \hat{k}} e^{2 i \hat{k} \ln{2 m \Omega r_0}} \ ,
  \end{equation}
  where $\Gamma$ is the Euler gamma function, $\hat{k} = m \Omega$ in the TM limit and $r_0 = 2/\sqrt{e}$.
  The remaining part of the factorized modes \eqref{fact_modes} is expressed as an amplitude $f_{\ell m}$ and a phase $\delta_{\ell m}$, 
  which are computed such that the expansion of $h_{\ell m}$ agrees with the PN expanded modes.
  We point the reader to Appendix B of Ref.~\cite{Pompili:2023tna} for the explicit expressions of the different $f_{\ell m}$ and $\delta_{\ell m}$ terms.
  
  The TM limit of Eqs.~\eqref{QC force} and ~\eqref{fact_modes} is obtained by setting the mass ratio $\nu$ to zero in the expressions, except for the leading $\nu$ in the Newtonian prefactor~\eqref{Newtprefactor}.

  In our analysis, we consider equatorial planar geodesics of the Kerr background; hence, we consider $\mF_{r} = \mF_{\varphi} = 0$ in Eqs.~\eqref{Ham_EOM_3} and \eqref{Ham_EOM_4}, when evolving the dynamics.
  We characterize the planar orbits through the parameters $\{p, e, a\}$, which correspond to the semilatus rectum, the eccentricity and the spin of the Kerr BH, respectively.
We adopt the \emph{Keplerian parametrization}, where for the definitions of $p$ and $e$ we have:
  \begin{equation}
    p = \frac{2r_{\text{a}}r_{\text{p}}}{r_{\text{a}} + r_{\text{p}}} \quad , \quad e = \frac{r_{\text{a}} - r_{\text{p}}}{r_{\text{p}} + r_{\text{a}}} \quad , 
  \end{equation}
  where $r_{\text{a}}$ and $r_{\text{p}}$ are the radial separation at the apocenter and at the pericenter, respectively.

  As we show in Sec.~\ref{Subsec: Analytical fluxes and Schott terms}, after evolving Eqs.~(\ref{Ham_EOM}) without RR forces (i.e., for geodesics), we compute the analytical fluxes by evaluating the RR force in Eqs.~\eqref{ecc impl} on the geodesics.
  Hence, to avoid any possible confusion to the reader we stress that whenever the QC expressions in Eqs.~\eqref{QC force} and~\eqref{fact_modes} are employed, they are evaluated on the geodesic although they are quantities constructed assuming QC trajectories.

  % ------------------------
  % Numerical fluxes section
  % ------------------------
  \subsection{Numerical fluxes} \label{Subsec: Numerical fluxes}
  The core of our analysis relies on the computation of the energy flux $ \Phi_{E}$ and angular-momentum flux $\Phi_{J}$
  radiated by the TM to future null infinity. These fluxes are computed numerically by solving the Teukolsky master equation~\cite{Teukolsky:1973ha}, which
  in Boyer-Lindquist coordinates reads
  \begin{equation} \label{Teukolsky equation}
    \begin{aligned}
      % & - \left[ \frac{(r^2 + a^2)^2}{\Delta} -a^2\sin{\theta}^2 \right]\partial_{tt}\Psi - \frac{4Mar}{\Delta} \partial_{t \varphi} \Psi \\
      % & - 2s \left[ r - \frac{M(r^2 - a^2)}{\Delta} + ia\cos{\theta} \right] \partial_{t} \Psi + \Delta^{-s} \partial_{r}(\Delta^{s+1} \partial_r \Psi) \\
      % & + \frac{1}{\sin{\theta}} \partial_{\theta} (\sin{\theta} \partial_{\theta} \Psi) + \left[ \frac{1}{\sin{\theta}^2} - \frac{a^2}{\Delta} \right]\partial_{\varphi \varphi} \Psi \\
      % & + 2s \left[ \frac{a(r - M)}{\Delta} + \frac{i\cos{\theta}}{\sin{\theta}^2} \right] \partial_{\varphi} \Psi - (s^2 \cot{\theta}^2 - s) \Psi \\
      % & = -4 \pi (r^2 + a^2 \cos{\theta}^2) T \ .
      & - \left[ \frac{(r^2 + a^2)^2}{\Delta} -a^2\sin{\theta}^2 \right]\partial_{tt}\Psi - \frac{4ar}{\Delta} \partial_{t \varphi} \Psi \\
      & - 2s \left[ r - \frac{(r^2 - a^2)}{\Delta} + ia\cos{\theta} \right] \partial_{t} \Psi + \Delta^{-s} \partial_{r}(\Delta^{s+1} \partial_r \Psi) \\
      & + \frac{1}{\sin{\theta}} \partial_{\theta} (\sin{\theta} \partial_{\theta} \Psi) + \left[ \frac{1}{\sin{\theta}^2} - \frac{a^2}{\Delta} \right]\partial_{\varphi \varphi} \Psi \\
      & + 2s \left[ \frac{a(r - 1)}{\Delta} + \frac{i\cos{\theta}}{\sin{\theta}^2} \right] \partial_{\varphi} \Psi - (s^2 \cot{\theta}^2 - s) \Psi \\
      & = -4 \pi (r^2 + a^2 \cos{\theta}^2) \mathcal{T} \ .
    \end{aligned}
  \end{equation}
  This equation describes the evolution of scalar, vector, and tensor perturbations of a Kerr BH.
  The function $\Delta$, the spin parameter $a$ and the coordinates correspond to the same quantities defined in the previous sections.
  The parameter $s$ is the \textit{spin weight} of the field. In particular, when $s = \pm 2$ the equation describes degrees of freedom of gravity that radiate, and for $s = -2$ it is $\Psi = (r - ia\cos{\theta})^4\psi_4$, where 
  $\psi_4$ is the Weyl curvature scalar that describes outgoing GWs.
  
  A system composed of a TM orbiting a Kerr BH can be interpreted as a perturbed Kerr metric, and within this interpretation the source term $\mathcal{T}$ in the right-hand side of Eq.~\eqref{Teukolsky equation}
  describes a TM moving in the Kerr spacetime. 
  The details on how the source term $\mathcal{T}$ of the TM is constructed and how Eq.~\eqref{Teukolsky equation} is numerically solved are beyond the scope of this section. 
  We mention the fact that the source term $\mathcal{T}$ of a TM orbiting a Kerr BH is constructed from Dirac-delta functions of the variables $r$ and $\theta$, as well as first and second derivatives of the delta functions in these variables. 
  These terms are sourced at the location of the TM, hence the source $\mathcal{T}$ depends on the trajectory that the TM follows in the Kerr spacetime. The details can be found in Refs.~\cite{Sundararajan:2008zm, Field:2020rjr, vandeMeent:2015lxa}.
  In this analysis, the trajectories used to source the term $\mathcal{T}$ are the geodesics introduced at the end of Sec.~\ref{Subsec:EOB model in the TM limit}, constructed evolving Eqs.~\eqref{Ham_EOM}.

  To solve Eq.~\eqref{Teukolsky equation}, we make use of two different codes:
  when considering bound orbits we adopt the FD code of Ref.~\cite{vandeMeent:2015lxa}, while when considering unbound orbits we employ the TD code developed in Ref.~\cite{Sundararajan:2007jg, Sundararajan:2008zm, Field:2020rjr}.
  At future null infinity, the Weyl scalar $\psi_4$ and the waveform strain $h = h_{+} - i h_{\times}$ are related by the expression
  \begin{equation}
    \psi_4 = \frac{1}{2}\frac{\partial^2 h}{\partial t^2} = \frac{1}{2} \left( \frac{\partial^2 h_{+}}{\partial t^2} - i\frac{\partial^2 h_{\times}}{\partial t^2} \right) \ ,
  \end{equation}
  and following standard practices, the waveform is decomposed in spin-weighted spherical harmonics
  \begin{equation}
    h = \sum_{\ell, m} {}_{-2}Y_{\ell m}(\theta, \varphi)\,h_{\ell m}  \ .
  \end{equation} 
  The fluxes $\Phi_{E}$ and $\Phi_{J}$ are computed from the modes through the expressions
  \begin{subequations}
  	\label{eq:inst_fluxes}
    \begin{align}
      & \Phi_{E} = \frac{1}{16 \pi} \sum_{\ell m} |\dot{h}_{\ell m}|^2 \label{E_flux_from_modes} \ , \\
      & \Phi_{J} = \frac{1}{16 \pi} \sum_{\ell m} m \Im{(\dot{h}_{\ell m} h_{\ell m}^{*})}  \label{J_flux_from_modes} \ .
    \end{align}
  \end{subequations} 
  In Appendix~\ref{App:FDInstFlux}, we give closed form expressions for constructing the instantaneous fluxes from FD Teukolsky solutions.
  In this work, we truncate the summation over the modes at $\ell = 8$. More details on the numerical errors of the FD and TD Teukolsky codes employed in our analysis can be found in Refs.~\cite{vandeMeent:2015lxa, Sundararajan:2007jg, Sundararajan:2008zm, Barausse:2011kb, Field:2020rjr}.

  % ----------------------------------------
  % Analytical fluxes and Schott terms
  % ----------------------------------------
  \subsection{Analytical fluxes} \label{Subsec: Analytical fluxes and Schott terms}
  In order to assess the EOB eccentric RR force of Eqs.~\eqref{ecc impl} by comparing with numerical results, it is necessary to compute the analytical fluxes.
  The connection between the RR force and the fluxes is given by the \emph{balance equations}
  \begin{subequations} 
  \label{eq:balance_equations}
  \begin{align}
    & \dot{E}_{\text{system}} + \Phi_{E}  + \dot{E}_{\text{Schott}}  = 0 \label{Energy balance equations} \ , \\ 
    & \dot{J}_{\text{system}} + \Phi_{J}  + \dot{J}_{\text{Schott}}  = 0 \label{Angular M. balance equations} \ .
  \end{align}
  \end{subequations} 
  These equations relate the time-dependent fluxes at future null infinity, $\Phi_{E}$ and $\Phi_{J}$, to the change in the energy 
  and angular momentum of the system, $\dot{E}_{\text{system}}$ and $\dot{J}_{\text{system}}$, together with two other terms that appear as total time derivatives, $\dot{E}_{\text{Schott}}$ and $\dot{J}_{\text{Schott}}$,
  known as Schott terms. These two terms take into account the contributions to the fluxes due to the interaction of the system with the radiation field, as originally pointed out in the context of electromagnetism in Ref.~\cite{Schott} 
  and they were introduced in the context of the EOB framework in Ref.~\cite{Bini:2012ji}. 
   
  The connection between the fluxes $\Phi_{E/J}$, the RR force $\mF$ and the Schott terms is made explicit by first considering the Hamilton equations~\eqref{Ham_EOM}, which lead to
  \begin{subequations}
  \label{eq:dotE_dotJ_system}
  \begin{align}
    & \dot{E}_{\text{system}} = \frac{d H}{dt} = \dot{r}\mF_{r} + \dot{\varphi}\mF_{\varphi} \ ,  \\
    & \dot{J}_{\text{system}} = \pphidot = \mF_{\varphi} \ ,
    \label{eq:dotJsystem}
  \end{align}
  \end{subequations}
  where the fact that the Hamiltonian in Eq.~\eqref{Eq: Kerr_Hamiltonian} does not depend on the azimuthal angle $\varphi$ is exploited.
  By plugging these expressions in Eqs.~\eqref{eq:balance_equations}, one gets:
  \begin{figure*}
  	\includegraphics[width=1.\linewidth]{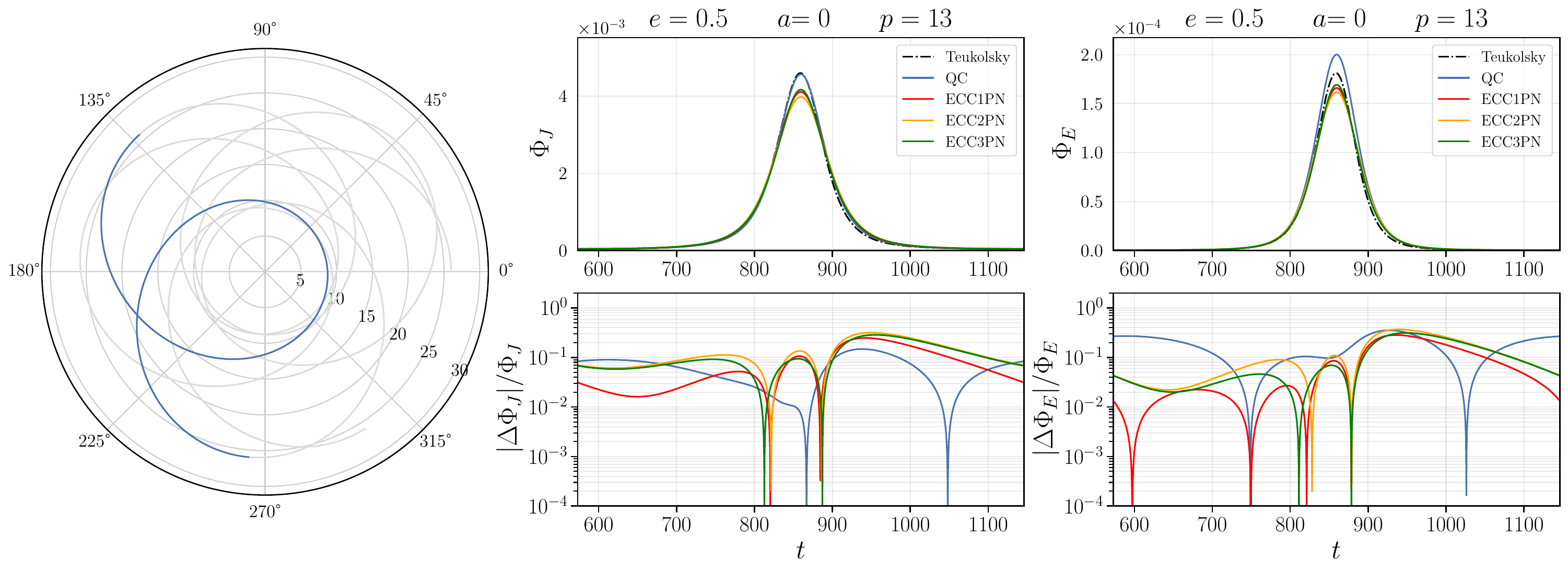}
  	\caption{Fluxes for an eccentric geodesic with $p = 13$, $e = 0.5$ and $a=0$. On the left panel,
  		the planar orbit is shown, highlighting in blue the trajectory over one radial period. 
  		The top panels show the angular momentum $\Phi_{J}$ and energy $\Phi_{E}$ fluxes at infinity:
  		black curves are the Teukolsky fluxes, while colored curves are the analytical fluxes. 
  		``QC" refer to the QC fluxes while ``ECCnPN'' refer to fluxes computed considering 
  		eccentric corrections at nPN order.
  		The bottom panels show the relative difference with respect to the Teukolsky fluxes.}
  	\label{inst_plot_1}
  \end{figure*}
  
  \begin{subequations} \label{Fluxes_RRforce_dependence}
  \begin{align}
    & \Phi_{E} = - \dot{r}\mF_{r} - \dot{\varphi}\mF_{\varphi} - \dot{E}_{\text{Schott}}\ , \label{E_flux_RRforce_dependence}\\
    & \Phi_{J} = - \mF_{\varphi} - \dot{J}_{\text{Schott}} \label{J_flux_RRforce_dependence} \ .
  \end{align}
  \end{subequations}
  From these equations we are able to compute the fluxes from the RR force in Eqs.~\eqref{ecc impl}, providing a way to test its prescriptions. This is done by first computing Kerr geodesics as explained in Sec.~\ref{Subsec:EOB model in the TM limit} 
  and then evaluating all the quantities involved in the right-hand side of Eqs.~\eqref{Fluxes_RRforce_dependence} on the geodesics.
  We point the reader to Appendix~\ref{Appendix A} for the expressions of the PN time derivatives of the Schott terms, $\dot{E}_{\text{Schott}}$ and $\dot{J}_{\text{Schott}}$, as functions of the EOB dynamical variables, as used in this work.In these expressions, which are written for the first time in this work, we follow the gauge choice employed in Ref.~\cite{Gamboa:2024imd} by fixing the gauge constants as $\alpha=-16/3$ and $\beta=-13/2$.

 	Note that in the above, we have not included any effects due to the central Kerr BH absorbing GWs, changing its mass and spin.
 	If included, these effects would alter the relationship between the RR forces $\mF_{r/\varphi}$ and fluxes $\Phi_{E/J}$ in Eqs.~\eqref{Fluxes_RRforce_dependence}.
 	In practice, the effects of absorption are typically several orders of magnitude smaller than the fluxes to infinity see e.g.~\cite{Drasco:2005kz} (but can become order $10\%$ for extreme orbits close to the horizon of a nearly extremal BH~\cite{Kapadia:2013kf,Taracchini:2013wfa,Colleoni:2015ena}).
 	As such, the absorption fluxes are mostly relevant when their effects can accumulate over a large number of orbits in an inspiral.
 	For the single orbit comparisons in this work, their impact would be minimal.
 	Nonetheless, to ensure an apples-to-apples comparison, we also include only the fluxes to infinity in the numerical Teukolsky fluxes.

  % ----------------------------------------
  % ------------- Results --------------
  % ----------------------------------------
\section{Comparison of the analytical and numerical fluxes} \label{Sec: Results}

In the following, we present the comparison between the analytical and numerical fluxes computed as explained in Sec.~\ref{Sec. II}.
In particular, we focus on studying the different PN contributions in the eccentric part of the analytical fluxes. 
Throughout this section, we label the fluxes that include PN corrections due to eccentricity at 1PN, 2PN and 3PN order (including SO and SS corrections up to 2PN order), as ``ECC1PN'', ``ECC2PN'' and ``ECC3PN'', while the fluxes that simply use the QC prescription of the fluxes evaluated on an eccentric trajectory are labeled ``QC''. Note that the fluxes labeled ``ECCnPN'' still contain all available PN orders in the QC part of the flux. 

% ------------- Instantaneous fluxes ------------- -------------
\subsection{Instantaneous fluxes} \label{Subsec: Instantaneous fluxes}

We start by comparing the numerical and analytical instantaneous fluxes. 
Before doing so, we emphasize that the Schott terms in Eqs.~\eqref{Fluxes_RRforce_dependence}, are known only as PN expansions up to 3PN order, while the RR force terms have been resummed in the EOB formalism. 
Consequently, there is a limit to what can be learned when comparing the instantaneous fluxes to the numerical-Teukolsky fluxes in the strong-field regime, 
as it is unclear whether any particular disagreement is due to an inaccuracy of the RR forces or a possible degradation of the PN approximation in the Schott terms. 
This ambiguity is not present when considering orbit-averaged fluxes, as we mention in Sec.~\ref{Subsec: Bound orbits in Schwarzschild}.
Nonetheless, we believe it is instructive to look at the comparison of the instantaneous fluxes as this helps build an intuitive picture of our analysis.

The comparison of the  numerical and analytical fluxes is illustrated in Fig.~\ref{inst_plot_1}. 
We consider an orbit with semilatus rectum $p = 13$, eccentricity $e = 0.5$ and spin $a = 0$.
The analytical and numerical fluxes are plotted over a radial period on the bound geodesic. In the left panel, the orbit is shown, while in
the top panels the instantaneous fluxes between two consecutive apocenters are plotted. The black curves correspond to the numerical 
fluxes obtained from the FD Teukolsky code, while the colored lines are the analytical fluxes. 
We show the QC fluxes (blue curves), which are computed with the QC RR force in Eqs.~\eqref{QC force}, and the eccentric fluxes ECC1PN (red curves), ECC2PN (orange curves) and ECC3PN (green curves).
Finally, the bottom panels of Fig.~\ref{inst_plot_1} show the fractional difference between the analytic and Teukolsky fluxes. 
The analytical fluxes are computed considering the multiplicative implementation \eqref{multiplicative impl}. The additive corrections provide similar results to Fig.~\ref{inst_plot_1}, and are not shown here.

For the geodesic considered in Fig.~\ref{inst_plot_1}, we do not find a clear improvement of the fluxes at different PN orders along the orbit.
Around the apocenter passage (for $t\approx 600 $ and $t\approx 1100 $ where $r \approx 25 $) we observe that the ECC1PN fluxes better approximate the numerical ones, 
whereas the ECC2PN and ECC3PN fluxes are close, but do not improve the approximation.
By removing the eccentric PN tail terms, we find that the 1.5PN tail contribution is degrading the accuracy of the instantaneous 
results. However, in Sec.~\ref{Subsec: Bound orbits in Schwarzschild}, we will discuss that these terms are essential to recover the correct scaling in the weak-gravitational field regime when considering the orbit-averaged fluxes.
This confirms what we anticipated at the beginning of this section: one PN order may be worse for the instantaneous fluxes, but better for the averaged fluxes, and thus, we cannot conclude from those 
results if the RR force is accurate since we cannot disentangle it from the effect of the PN-expanded Schott terms.

Near the pericenter passage (for $ t\approx 850 $ where $r \approx 9 $) we find that the ECC3PN energy flux improves the other PN orders, while 
for the angular-momentum flux it is the QC curve that best approximates the numerical flux. 
This close agreement of the QC fluxes near the pericenter passage was already pointed out in Ref.~\cite{Albanesi:2022ywx} for some orbital configurations, and explained as a numerical coincidence.
This will become more apparent in Sec.~\ref{Subsec: Bound orbits in Schwarzschild}, as we show the behavior of this agreement as a function of the orbital separation.

The bottom panels of Fig.~\ref{inst_plot_1} show an asymmetry of the relative differences with respect to the pericenter passage, which was also observed in Ref.~\cite{Albanesi:2021rby}.
After convincing ourselves that this behavior is not due to any numerical artifact, we find that it comes from an asymmetry of the Teukolsky fluxes with respect to the pericenter. 
We conclude that this asymmetry arises from contributions that are not modeled by the EOB fluxes, but that are present in the numerical fluxes.
By inspecting different orbits and observing the same pattern, especially for orbits that lay in the weak-field regime, we suggest that this asymmetry comes from delayed contributions coming from higher-order-PN tail terms that are not present in the analytical fluxes. 

We notice that there are qualitative differences between our results and Fig.~1 of Ref.~\cite{Albanesi:2022ywx}, despite the two figures ostensibly plotting the same quantities for the same orbit with both the QC and ECC2PN (labeled ``QC2PN'' in Ref.~\cite{Albanesi:2022ywx}); in our version, the fluxes are much closer to the numerical values.
The discrepancy in the QC flux may arise due to the proxy used in Ref.~\cite{Albanesi:2022ywx} for the QC flux not faithfully reproducing the QC flux of \texttt{SEOBNRv4HM} (and notably the QC flux of \texttt{SEOBNRv5HM}, which is used in this paper) when applied to eccentric orbits. 
The discrepancy between the ECC2PN fluxes, on the other hand, is due to the fact that in Ref.~\cite{Albanesi:2022ywx}, the authors use the Schott terms of Ref.~\cite{Bini:2012ji} when computing the fluxes through Eqs.~\eqref{Fluxes_RRforce_dependence}.
However, this is not compatible with the gauge of the eccentric RR force of Ref.~\cite{Khalil:2021txt}, which is used to compute the ECC2PN (QC2PN) fluxes. There is no freedom in using Schott terms in a different gauge, and the ones of Ref.~\cite{Khalil:2021txt} must be considered when computing the instantaneous ECC2PN fluxes.

% ------------- Schwarzschild bound orbits --------------
\subsection{Averaged fluxes: bound orbits in Schwarzschild} \label{Subsec: Bound orbits in Schwarzschild} 
\begin{figure*} 
  \includegraphics[width=1.\linewidth]{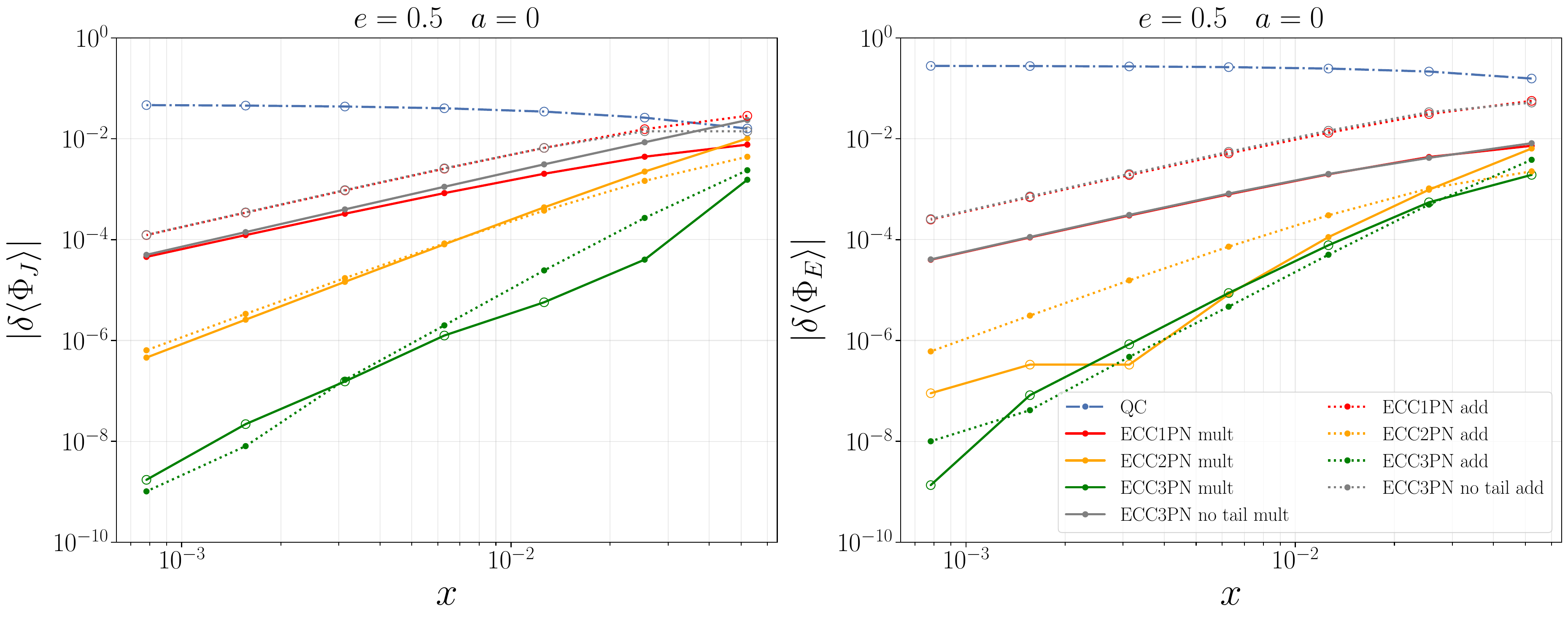}
  \caption{Absolute value of the fractional differences defined in Eq.~\eqref{Eq: fractional difference definition} of the averaged fluxes computed for orbits in the weak-field regime. The x-axis corresponds to the gauge-invariant variable $x = \langle \dot{\varphi} \rangle^{2/3}$. The orbits are characterized by the parameters $e = 0.5$, $a = 0$ and $p = \{ 960, 480, 240, 120, 60, 30, 15 \}$. Dotted lines correspond to the additive implementation (add) while solid ones correspond to the multiplicative (mult). Empty circle dots represent negative values of the fractional differences, highlighting zero crossing in the log-log plots. }
  \label{Fig: averaged-fluxes WF}
\end{figure*}
Figure~\ref{inst_plot_1} indicated that the 1PN eccentric corrections of the multiplicative implementation 
of the RR force~\eqref{multiplicative impl} better approximate the numerical flux than the higher order eccentric PN corrections  near the pericenter of the orbit. 
The additive implementation~\eqref{additive impl} (not plotted in Fig.~\ref{inst_plot_1}) shows similar behavior.
This motivates a further investigation, because, in principle, one would expect the higher order PN corrections to improve the approximation, at least in the  weak-field regime.

To better assess this, we start by restricting to the Schwarzschild case and we analyze what happens when considering the eccentric force in the weak-field scenario.
For the study, we consider the averaged fluxes over one radial period $T_r$, given by:
\begin{equation} \label{avg_fluxes_def}
  \langle \Phi_{E,J} \rangle = \frac{1}{T_r}\int_{t_0}^{t_0+T_r}{dt \; \Phi_{E,J}} \ \, .
\end{equation}
The averaging eliminates the ambiguity due to the Schott terms contribution in the instantaneous-fluxes expressions \eqref{Fluxes_RRforce_dependence}, 
as the Schott terms appear as total time derivatives.
 
We consider a set of bound orbits defined by eccentricity $e=0.5$, spin $a=0$ and decreasing semilatera recta $p = \{ 960, 480, 240, 120, 60, 30, 15 \}$.
This choice allows us to explore the weak-field regime in order to check whether the flux residuals, obtained by subtracting the eccentric corrections \eqref{multiplicative impl} and \eqref{additive impl} from the numerical fluxes, have the expected PN scaling. 

In Fig.~\ref{Fig: averaged-fluxes WF}, we show the fractional difference of the averaged numerical fluxes with respect to the averaged analytical fluxes
\begin{equation} \label{Eq: fractional difference definition}
    \delta \langle \Phi_{E,J} \rangle = \frac{\langle \Phi_{E,J}^{\rm numerical} \rangle - \langle \Phi_{E,J}^{\rm analytical} \rangle}{\langle \Phi_{E,J}^{\rm numerical} \rangle}
\end{equation}
computed on these orbits for different values of the gauge-invariant quantity $x$ defined by
\begin{equation} \label{x_def}
x = \langle \dot{\varphi} \rangle^{2/3} \ .
\end{equation}
For the different PN corrections, we follow the same nomenclature used in Fig.~\ref{inst_plot_1}, and we consider both implementations: solid lines correspond to the multiplicative RR force from Eq.~\eqref{multiplicative impl} while dotted lines correspond to the additive RR force from Eq.~\eqref{additive impl}.

We observe that the eccentric corrections to the RR force provide a consistent improvement over the QC force. For orbits that are in the weakest regimes, $x \approx 8\times 10^{-4}$, including the ECC3PN corrections improves the agreement with the numerical flux by a factor $~10^{8}$ over the QC flux.
We also find that the curves follow the general expected scaling at low $x$.
The fractional difference between the QC analytical and the Teukolsky flux approaches a constant in the weak field regime, indicating that the QC prescription already needs corrections at the leading ``Newtonian'' order.
The slopes of fractional residuals, after subtracting the ECCnPN corrections in the log-log plot of Fig.~\ref{Fig: averaged-fluxes WF}, are compatible with the expected $x^{-(n+1/2)}$ behavior of a residual that starts at $(n+1/2)$PN order.

Since for the instantaneous flux comparison in Sec.~\ref{Subsec: Instantaneous fluxes} the inclusion of the PN tail corrections did not lead to an unequivocal improvement of the flux, we also considered the effects of omitting the PN tail terms on the orbit-averaged flux.
The gray lines in Fig.~\ref{Fig: averaged-fluxes WF} correspond to the ECC3PN fluxes computed without the tail part, showing that the inclusion of these terms is essential in order to recover the expected PN scaling, 
corroborating the hypothesis that degradations of higher-order terms in the instantaneous fluxes may come from the Schott terms and disappear when orbit averaging.

When moving into the stronger-field regimes (i.e. for higher values of $x$), the discrepancy between analytical and numerical averaged fluxes increases, as expected due to the PN nature of the analytical fluxes. 
\begin{figure*} 
  \includegraphics[width=1.\linewidth]{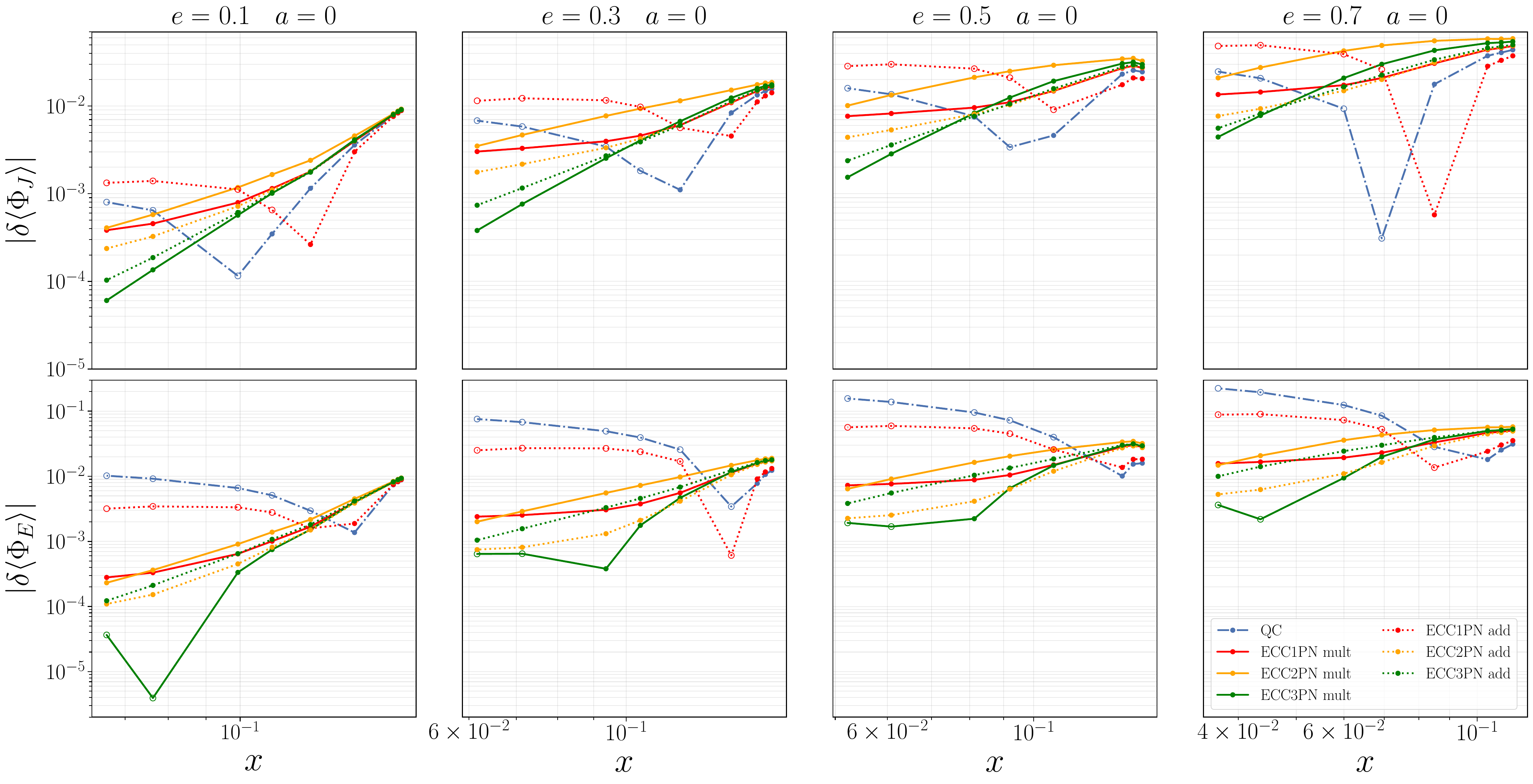}
  \caption{Absolute value of the fractional differences of the averaged fluxes computed on orbits in mild/strong-field regimes. The orbits with highest $x$ (last points on the right) are characterized by a semilatus rectum $p = p_{\rm LSO} + 0.025$. The nomenclature is the same as in Fig.~\ref{Fig: averaged-fluxes WF}. } 
  \label{Fig: averaged-fluxes SF}
\end{figure*}

In order to assess the RR force over a wider region of the parameter space, we consider more orbits with different eccentricities spanning the milder and stronger field regimes. 
~Figure~\ref{Fig: averaged-fluxes SF} shows the fractional differences of the fluxes evaluated at geodesics with eccentricities $e=\{0.1, 0.3, 0.5, 0.7\}$, spin $a=0$ and semilatera recta $p = \{ p_{\rm LSO}+0.025, p_{\rm LSO}+0.05, p_{\rm LSO}+0.1, 7, 8, 9, 10, 13, 15 \}$ \footnote{
For the orbits with $e=\{ 0.5, 0.7 \}$, we consider the set $p = \{ p_{\rm LSO}+0.025, p_{\rm LSO}+0.05, p_{\rm LSO}+0.1, 8, 9, 10, 13, 15 \}$, since $p_{\rm LSO} \ge 7$ for these orbits.
We also limit the eccentricity to $\leq 0.7$ since frequency-domain Teukolsky codes become computationally expensive for high eccentricities.
}, where $p_{\rm LSO}$ corresponds to the 
semilatus rectum of the last stable orbit (LSO) with the corresponding $e$. 
For Schwarzschild geodesics this is given by
\begin{equation}
  p_{\rm LSO} = 6+2e \ .
\end{equation}

The strong-field residuals in Fig.~\ref{Fig: averaged-fluxes SF} show a much less organized picture than their weak-field counterparts in Fig.~\ref{Fig: averaged-fluxes WF}. There is not always a clear order-by-order improvement from adding higher PN eccentric corrections. This signifies (the start of) the breakdown of the convergence of the PN series in this regime. Notably, the QC and ECC1PN fluxes seemingly outperform the higher-order corrections for larger $x$.
In fact---contrary to naive expectation---the residuals from the QC and ECC1PN fluxes actually decrease, for both the angular-momentum and energy fluxes, up to certain values of $x$.
This behavior is evident starting from values of $x \approx 0.06$ for the orbits with $e = \{ 0.1, 0.3, 0.5 \}$ and of $x \approx 0.04$ for the ones with $e = 0.7$.
This decreasing trend is interrupted at larger values of $x$, where the QC and ECC1PN residuals of the fluxes start to monotonically increase up to the
closest orbits to the LSO with $p = p_{\rm LSO} + 0.025$.
We find that this is connected to a change in sign of the fractional differences between the analytical/numerical 
fluxes. 
Since the absolute fractional differences are plotted in Fig.~\ref{Fig: averaged-fluxes SF}, here we represent 
this change of sign by using different dots: filled dots for positive values of the relative difference~\eqref{Eq: fractional difference definition}, empty dots for negative values.
This change of sign signifies that in the weak field the average QC fluxes overestimate
the average numerical flux, while they underestimate it in the stronger field.
Consequently, there are some ``goldilocks'' configurations for which the QC (and ECC1PN) fluxes are ``just right'', and by coincidence produce the correct flux, corroborating what was found in Ref.~\cite{Albanesi:2022xge} for the QC fluxes.

\begin{figure*}[!tb]
	\includegraphics[width=1.\linewidth]{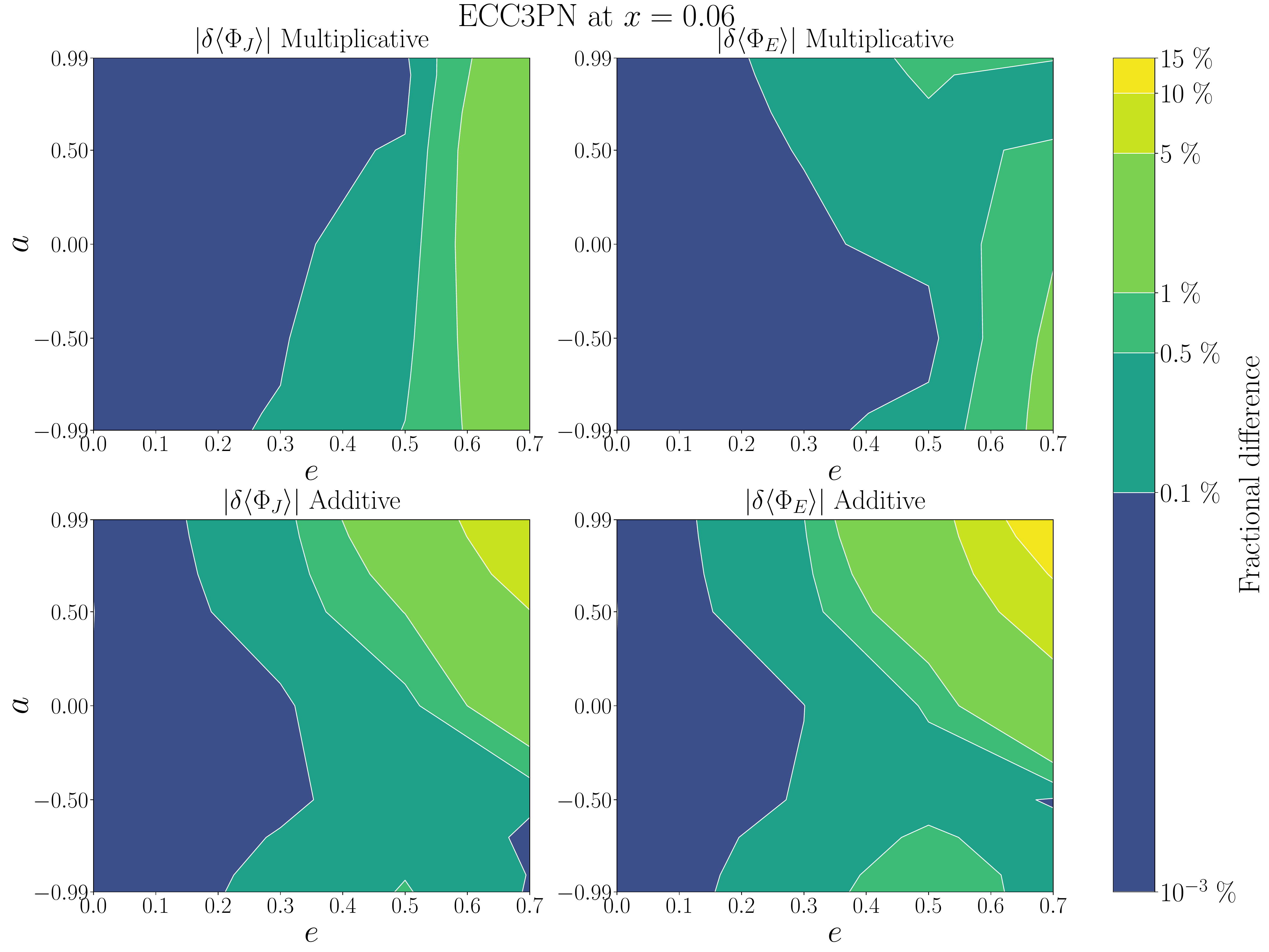}
	\caption{Contour plots of the absolute value of the fractional differences of the ECC3PN averaged fluxes with respect to the numerical one. They are computed on orbits that span the parameter space ($e$, $a$) with fixed value $x = 0.06$. The upper panels show the multiplicative implementation while the lower ones 
		show the additive implementation. The multiplicative implementation shows a worst case scenario of $5 \%$ relative difference for higher eccentric spinning prograde orbits.}
	\label{spin and ecc}
\end{figure*}

We do not observe a significant difference between the multiplicative and additive implementations in the non-spinning case. However, in Sec.~\ref{spinning orbits} we show that this changes when considering spin.

We also find that larger eccentricity values impact the accuracy of the analytical averaged fluxes. More specifically, we observe that for similar values of the $x$ parameter, the relative differences are more prominent for higher eccentricities. 
This trend is due to the fact that the pericenter of the orbits---where most radiation is generated---is pushed more and more in the strong-field regime, where in turn, the PN expressions are less reliable.

Figure~\ref{Fig: averaged-fluxes SF} allows us to gauge the overall performance of the flux approximations: in the worst case scenario (i.e. for orbits with $e=0.7$ and close to the LSO), we observe a difference of $\approx 5 \%$, when considering the ECC3PN fluxes.

% ------------- Kerr bound orbits --------------
\subsection{Averaged fluxes: bound orbits in Kerr} \label{spinning orbits}

So far, we have shown a comparison between analytical and numerical fluxes for eccentric (bound) orbits of a TM moving on geodesics around a Schwarzschild BH. In this section, we provide results 
that assess the RR force corrections in Eqs.~\eqref{ecc impl} when we allow the central BH to have a spin aligned or antialigned (henceforth, for simplicity aligned) with the orbital angular momentum (i.e., we consider eccentric-equatorial orbits around a Kerr BH).

To test the spinning case, we consider orbits with a fixed value of the $x$ parameter defined in Eq. \eqref{x_def}. This choice allows us to identify in 
a gauge-invariant manner different gravitational-field regimes. 
The orbits we analyze have eccentricity values $e = \{0.0, 0.1, 0.3, 0.5, 0.7 \}$ 
and spins of the central BH $a = \{ -0.99,-0.9 ,-0.7, -0.5, 0, 0.5, 0.7, 0.9, 0.99 \}$. 

In Fig.~\ref{spin and ecc}, we show contour plots of the fractional difference of the averaged fluxes \eqref{avg_fluxes_def} with respect to the eccentricity and the spin values for orbits with fixed $x = 0.06$. 
We choose this value for $x$ since it allows computing stable geodesics all over the subset of the parameter space $\{ e, a \}$ we want to investigate.
We consider the 3PN order in the eccentric sector, ECC3PN, for both implementations \eqref{multiplicative impl} and \eqref{additive impl}. 

We observe that, regardless of the spin values, for small eccentricities ($e \le 0.3$), the fractional difference is less than $0.5 \%$ for both resummations.
When considering higher values of $e$, the differences increase. 
This trend corroborates what we mentioned at the end of Sec.~\ref{Subsec: Bound orbits in Schwarzschild}, notably, for fixed values of $x$, larger eccentricities degrade the accuracy of the analytical-averaged fluxes, because the pericenter of the orbit is pushed more in the strong-field regime.
This degradation increases when considering prograde orbits with high-spin values, for which the pericenter is even closer to the central BH.
For these orbits, we find that the multiplicative implementation provides a mismatch of the fluxes less than $5 \%$, for the considered portion of parameter space. In comparison, the additive implementation provides relative differences that are less than $15 \%$. In particular, we find that for more extreme orbits (i.e., for the ones with high eccentricity and spin), 
the multiplicative implementation improves over the additive one by a $10 \%$ difference.

As a final remark, we mention an overall general improvement of the multiplicative ECC3PN factorization with respect to the additive ECC3PN factorization for regimes with $x \le 0.06$ over the parameter space $(e, a)$ for both the energy and angular-momentum fluxes. This can be seen in Fig.~\ref{Fig: averaged-fluxes-spinning WF} in Appendix~\ref{PN scaling assessment for spinning orbits} where
we show the behavior of the fluxes at different PN orders, for both factorizations, from weak-field regimes up to $x = 0.06$.
The considered orbits use the particular configuration of eccentricity and spin of Fig.~\ref{spin and ecc}, $e = 0.5$ and prograde spin $a = 0.9$, with semilatera recta $ p = \{ 960, 480, 240, 120, 60, 30, 15 \}$.
Figure~\ref{Fig: averaged-fluxes-spinning WF} also highlights (as in Fig.~\ref{Fig: averaged-fluxes WF}) the convergence of the different PN truncations of the eccentric fluxes 
pointing out the correctness of the procedure (and of the expressions) used to derive the EOB eccentric RR force in Eq.~\eqref{ecc impl} in the general spinning scenario.
We performed this same test for other eccentricity and spin configurations $(e, a)$ of Fig.~\ref{spin and ecc}, finding similar results.

% ------------- Schwarzschild unbound orbits --------------

\subsection{Scattering orbits in Schwarzschild} \label{Unbound orbits}
\begin{figure} 
	\includegraphics[width=1.\linewidth]{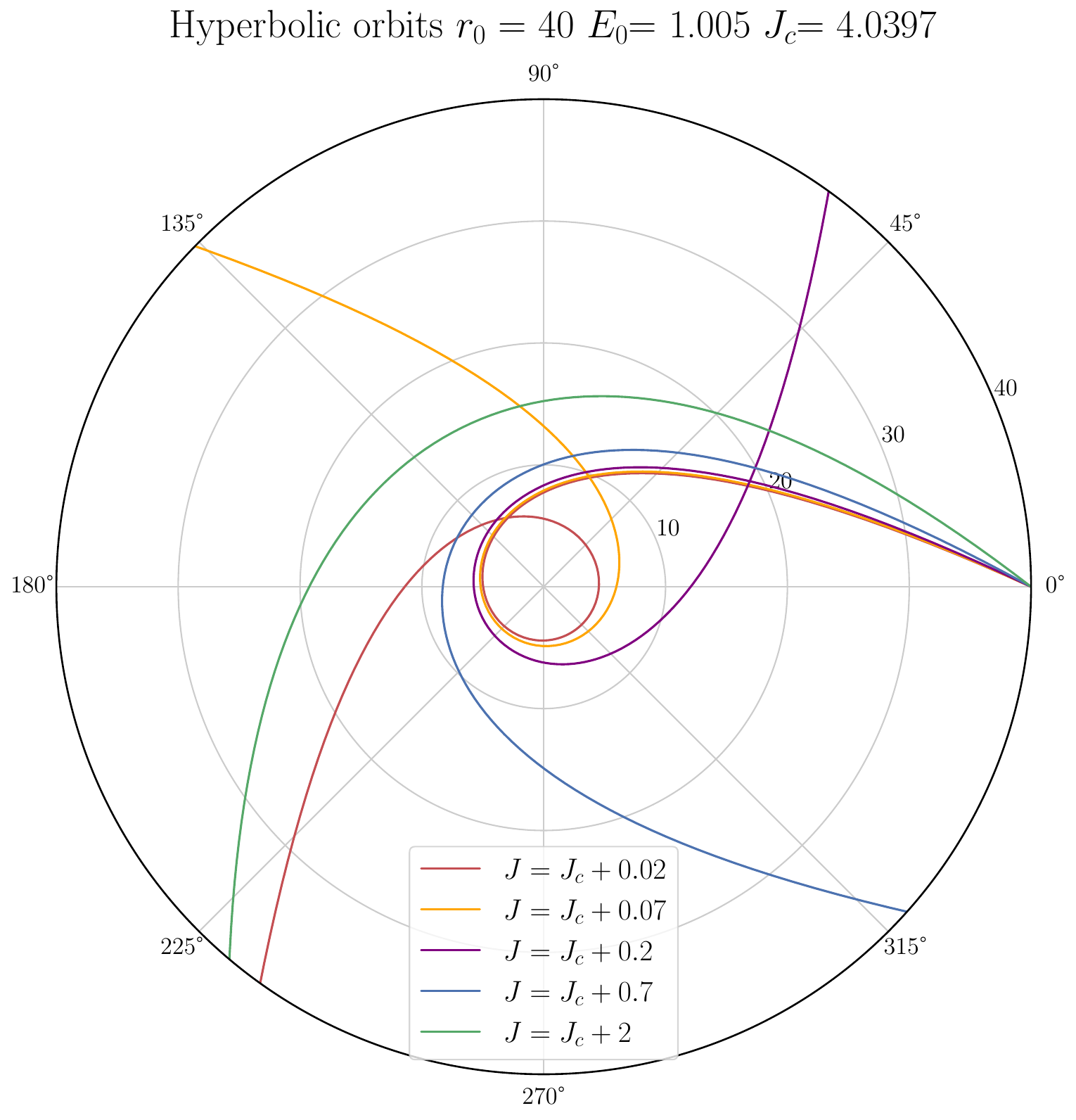}
	\caption{The hyperbolic orbits of a TM around a Schwarzschild BH considered in our analysis. They all have energy $E_0 = 1.005$ and critical angular momentum $J_c = 4.0397$. We consider five orbits which have total angular momenta $J_0 = \{J_c + 0.02, J_c + 0.07, J_c + 0.2, J_c + 0.7, J_c + 2 \}$. }
	\label{Considered hyperbolic orbits}
\end{figure}
\begin{table*}[th]
\setlength\tabcolsep{4pt}
	\centering
	\caption{Comparison of the analytical/numerical total emitted energy and angular momentum for the different hyperbolic orbits shown in Fig.~\ref{Considered hyperbolic orbits}. 
		The analytical fluxes are computed with the multiplicative implementation \eqref{multiplicative impl}. For each orbit with energy $E_0$ and angular momentum $J_0$, 
		we show the values of the pericenter distance $r_p$, the numerical total emitted energy $E_{\text{Teuk}}$ and angular-momentum $J_{\text{Teuk}}$, and the fractional 
		differences between the analytical emitted energy/angular-momentum, truncated at different PN orders in the eccentric sector, and the total numerical fluxes. } 
	\label{Hyper_flux_table}
\begin{ruledtabular}
	\begin{tabular}{ccc||c|cccc|c|cccc}
    $E_0 $ &  $J_0 $ & $r_{p} $ & $E_{\text{Teuk}}$ &  $\delta E_{\rm QC}$ &  $\delta E_{\rm 1PN}$ &  $\delta E_{\rm 2PN}$ &  $\delta E_{\rm 3PN}$ & $J_{\text{Teuk}}$ &  $\delta J_{\rm QC}$ &  $\delta J_{\rm 1PN}$ &  $\delta J_{\rm 2PN}$ &  $\delta J_{\rm 3PN}$ \\
		\midrule
		1.005 & 4.0597 & 4.39 & $2.605\times 10^{-1}$ & 0.3304 & 0.2095 & 0.2122 & 0.1330 & 2.805 & 0.2121 & 0.1948 & 0.1637 & 0.0998 \\
		1.005 & 4.1097 & 4.83  & $1.450\times 10^{-1}$ & 0.3316 & 0.1571 & 0.1650 &  0.0964 & $1.818 $ & 0.1701 & 0.1433 & 0.1135 & 0.0593 \\
		1.005 & 4.2397 & 5.64 & $6.680\times 10^{-2}$ & 0.3523 & 0.1056 & 0.1204 & 0.0807 & $1.062 $ & 0.1318 & 0.0927 & 0.0711 & 0.0412 \\
		1.005 & 4.7397 & 8.14 &  $1.381\times 10^{-2}$ & 0.4282 & 0.0473 & 0.0673 & 0.0631 & $3.836\times 10^{-1}$ & 0.0964 & 0.0394 & 0.0357 & 0.0370 \\
		1.005 & 6.0397 & 14.67 & $1.477\times 10^{-3}$ & 0.5225 & 0.0065 & 0.0093 & 0.0103 & $9.761\times 10^{-2}$ & 0.0911 & 0.0159 & 0.0224 & 0.0280 \\
	\end{tabular}
\end{ruledtabular}
\end{table*}
\begin{figure*} 
	\includegraphics[width=1.\linewidth]{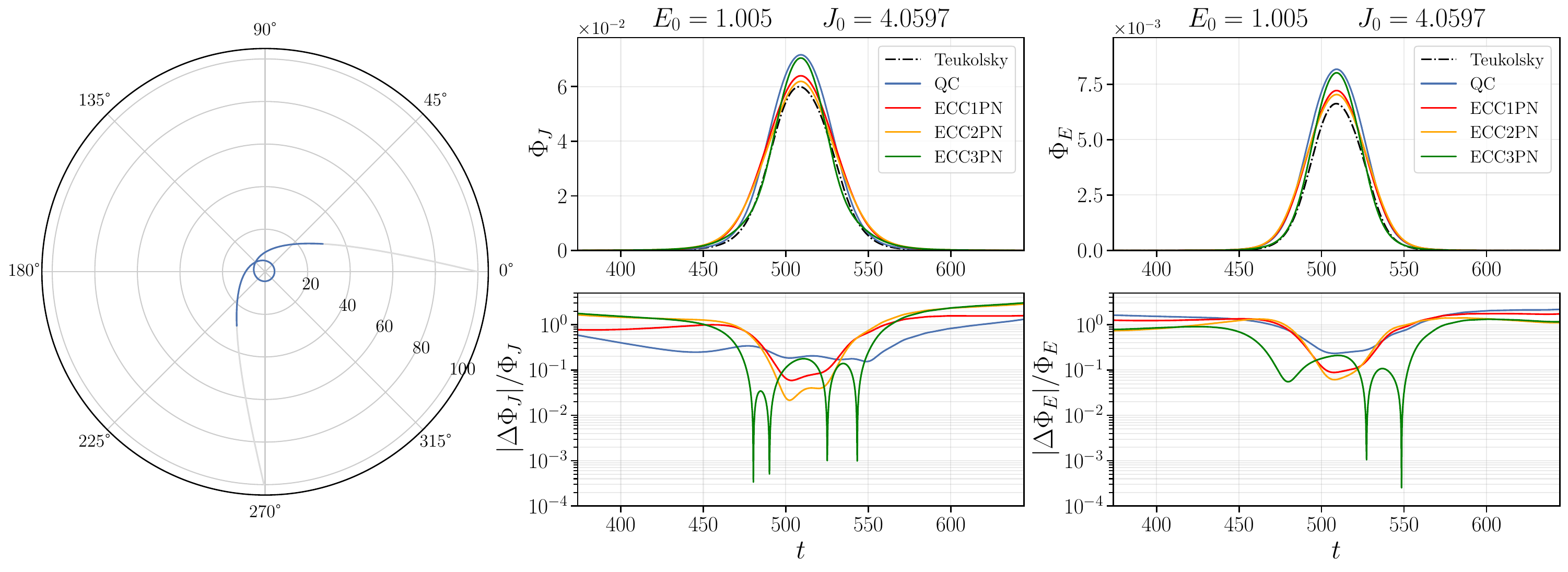}
	\includegraphics[width=1.\linewidth]{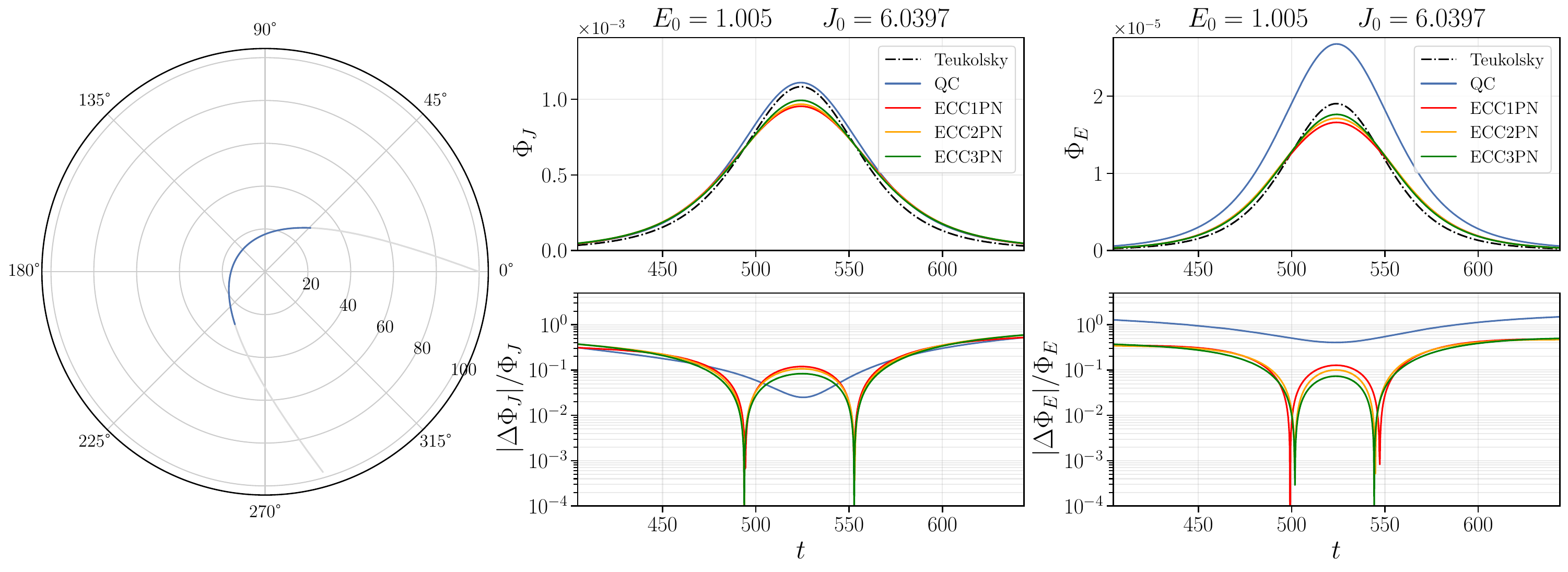}
	\caption{Instantaneous fluxes for hyperbolic orbits with $E_0 = 1.005$ and $J_0 = \{4.0597, 6.0397 \}$. Black curves are the numerical fluxes, blue-dotted ones represent QC fluxes, red curves are the fluxes containing 1PN eccentric corrections while
		orange and green curves are the eccentric 2PN and 3PN corrected fluxes. The portion of orbit we consider is highlighted in light-blue in the orbit-plot and it is for $r<30$.}
	\label{hyper_fluxes_in_time}
\end{figure*}

Finally, we push the comparison of the eccentric analytical and numerical fluxes to even higher eccentricities to the $e>1$ regime (i.e., hyperbolic-scattering orbits). In this section, we restrict our attention to non-spinning Schwarzschild BHs.

We consider hyperbolic orbits with fixed energy $E_0 = 1.005$, and we vary the angular momentum $J_0$. Figure~\ref{Considered hyperbolic orbits} shows the five different orbits we consider. 
To produce them, we examine different values of the angular momenta by first computing the critical value $J_c$ through Eq. (6) of Ref.~\cite{Barack:2022pde}. 
This critical angular momentum represents the smallest value a TM with fixed energy must have to still scatter back to infinity without plunging into the central BH.
For the chosen value of $E_0$, it is $J_c = 4.0397$. We consider different values for $J_0 = \{J_c + 2, J_c + 0.7, J_c + 0.2, J_c + 0.07, J_c + 0.02 \}$, which provide orbits with increasing scattering angle and decreasing pericenter distance. 
As in the bound case, we compute the analytical and numerical fluxes on these orbits. The numerical fluxes are obtained through the TD Teukolsky code of Refs.~\cite{Sundararajan:2007jg, Sundararajan:2008zm, Field:2020rjr} as mentioned in Sec.~\ref{Subsec: Numerical fluxes},
while the analytical fluxes are computed similarly to the bound-orbits case (i.e., by evaluating Eqs.~\eqref{Fluxes_RRforce_dependence} on the hyperbolic geodesics). 
In the following, we consider only the multiplicative implementation~\eqref{multiplicative impl}, since in Fig.~\ref{spin and ecc} we observe it improves over the additive one.

In Table~\ref{Hyper_flux_table}, we compare the analytical/numerical total energy and angular momentum emitted by the TM on the examined orbits. These quantities are evaluated by integrating the fluxes on each orbit and truncating at different PN orders.
As in the case of the averaged fluxes for bound orbits (see Sec.~\ref{Subsec: Bound orbits in Schwarzschild}), we also find that the eccentric corrections improve the agreement with the numerical fluxes with respect to the QC case.
We observe an improvement of the PN series in the eccentric sector, especially for orbits with smaller $J_0$. In almost all the examined cases, the 3PN corrections provide the best approximation of the total radiated energy and angular momentum with respect to the numerical flux. 
However, we find that the fractional differences for the orbits with smaller $J_0$ ($J_ 0 = 4.0597$ and $J_ 0 = 4.1097$) are
greater than $10\%$. This is expected, because these orbits have angular momentum close to the critical one $J_c$ and are characterized by strong-field regimes, where the PN expansions start to breakdown.
For the orbits with larger angular momentum, characterized by weaker-field regimes, the residuals are smaller ($\le 8 \%$) but we do not observe a large improvement of the PN series when increasing the order in general. Moreover, for the trajectory with higher angular momentum ($J_0 = 6.0397$), we observe a slight deterioration of the PN series,
with the eccentric 1PN corrections providing the best approximation.
We believe this is because even in the far weak field these orbits still have high (relativistic) velocities.

To provide an intuitive picture of this last point we consider again the instantaneous fluxes. Figure~\ref{hyper_fluxes_in_time} shows the time-dependent fluxes for the hyperbolic encounters with the highest ($J_0 = 4.0597$) and smallest ($J_0 = 6.0397$) scattering angle. From the plots, we observe that
for the orbit with the highest scattering angle, in the mild/weak-field regime part of the orbits (when $r>20$), the analytical fluxes differ by $>80 \%$ from the numerical ones.
This is particularly evident for the ECC2PN and ECC3PN fluxes, for which the fractional difference is $> 100 \%$.
We observe a similar pattern for the orbit with the smallest scattering angle ($J_0 = 6.0397$) (i.e., in the weak-field regime), a large mismatch between the numerical and the analytical curves arises.
This large fractional difference we observe for $r>20$ is caused by the fact that, for hyperbolic orbits, considering the weak-gravitational-field regime does not necessarily correspond to considering 
small velocities ($v \ll 1$) of the TM.
This large-velocity regime deteriorates the convergence of the PN expansion, as shown in the last line of Table~\ref{Hyper_flux_table}.

% ----------------------------------------
% ------------- Conclusions --------------
% ----------------------------------------

\section{Conclusions} \label{Conclusions}
In this work, we tested an EOB eccentric RR force computed with the same procedure as Ref.~\cite{Khalil:2021txt} with a different gauge choice for the leading PN order. The non-spinning 3PN part of this RR force is derived in Ref.~\cite{Gamboa:2024imd} while the 
SO and SS contributions, at 1.5PN and 2PN respectively, are similar to the ones of Ref.~\cite{Khalil:2021txt} but computed taking into account the different gauge choice of the leading order. 
We considered two possible resummations of this force: a multiplicative~\eqref{multiplicative impl} and an additive~\eqref{additive impl} one.
The assessment of the resummations is performed by considering a TM orbiting the equatorial plane of a Kerr BH. 
We computed the analytical energy and angular-momentum fluxes, which are linked to the RR force through the balance equations~\eqref{Fluxes_RRforce_dependence}, 
on eccentric geodesics of the Kerr metric, and we compared them with the numerical fluxes computed through the use of a FD~\cite{vandeMeent:2015lxa} and a TD~\cite{Sundararajan:2007jg,Sundararajan:2008zm,Field:2020rjr} Teukolsky codes.
We focused our analysis on the orbit-averaged fluxes for bound orbits and on the emitted energy and angular momentum for hyperbolic orbits. 
This is because when integrating the fluxes over an orbit, we are sure that any degradation coming from the PN-expanded Schott terms is integrated out and we can test the RR force resummations.

When considering bound orbits in the weak-field limit, we recovered the expected scaling of the different PN truncations of the fractional difference of the averaged fluxes, indicating that the procedure used to compute the RR force and its resummed expressions are both correct. 
We also showed that the eccentric corrections to the RR force are necessary to improve the QC fluxes in this regime. 
This improvement over the QC fluxes is of the order of $10^8$ when considering the ECC3PN fluxes for orbits with $x\approx10^{-3}$.
For non-spinning bound orbits in stronger-field regimes, computed up to the LSO, we found that the fractional differences of the averaged fluxes are $<5\%$ for eccentricity up to $e=0.7$. 
This result holds for both the multiplicative and additive resummations.

For the more general case of bound equatorial geodesics of a Kerr BH, characterized by $x \le 0.06$, $e = [ 0, 0.7 ]$ and $a = [ -0.99, 0.99 ]$, we found that for $e<0.3$ the fluxes discrepancies are $<0.5 \%$ for both resummations.
For larger eccentricities up to $e = 0.7$, we showed that the multiplicative implementation provides discrepancies with the numerical fluxes that are $<5\%$, while for the additive implementation, we found larger fractional differences, up to $\approx 15\%$ values. 
This indicates that, in the considered part of the parameter space, the multiplicative resummation~\eqref{multiplicative impl} is a better approximation of the fluxes up to regimes with $x = 0.06$. This is especially true for highly-eccentric ($e = 0.7$) and prograde orbits with large Kerr spin ($a = 0.99$).

Finally, we concluded our analysis by considering geodesic hyperbolic orbits with fixed energy $E = 1.005$ and different angular momenta. We found that the 3PN eccentric corrections of the multiplicative resummation improve the total emitted energy and angular momentum
for hyperbolic encounters with angular momenta closer to the critical one. However, for these orbits, the relative differences at 3PN are larger than $8 \%$, 
due to the stronger-field regimes characterizing them.
For the unbound orbits in weaker-field regimes, we found smaller values of the fractional differences ($< 7 \%$) of the emitted energy and angular momentum. However, we observed a deterioration of the convergence of the PN series, with the 3PN corrections providing higher discrepancies.
We pointed out this may be due to the fact that for unbound orbits the weak--gravitational-field regime does not necessarily imply small velocities, impacting the convergence of the eccentric PN orders.

The results we found for the unbound case indicate
that the RR force should be further improved for large
eccentricities and high velocities. There
are possible different strategies to tackle this last point: 
one possibility is to determine whether different parametrizations improve the results we find. 
Another strategy could be to perform a further resummation of 
the eccentric corrections to the RR force $F^{\rm ecc}_{r/\varphi}$ in Eqs.~\eqref{ecc impl}. 
A different approach would be to numerically inform the eccentric corrections by
introducing terms that fit the numerical fluxes in the parameter space. 
We believe all these strategies are promising for improving the EOB eccentric fluxes we studied in
this work, and we leave them to future works.
% ----------------------------------------
% ----------- Acknowledgments ------------
% ----------------------------------------
 
\section*{Acknowledgments}
The authors are grateful to Sergei Ossokine, Lorenzo Pompili, and Antoni Ramos-Buades for useful comments and suggestions. G.F. is grateful to D.M. for everything.
MvdM acknowledges financial support by the VILLUM Foundation (grant no. VIL37766) and the DNRF Chair program (grant no. DNRF162) by the Danish National Research Foundation. 
M.K.'s work is supported by Perimeter Institute for Theoretical Physics. Research at Perimeter Institute is supported in part by the Government of Canada through the Department of Innovation, Science and Economic Development and by the Province of Ontario through the Ministry of Colleges and Universities.
G.K. acknowledges support from NSF Grants No. PHY-2307236 and DMS-2309609. Some computations were performed on the UMass-URI UNITY HPC/AI cluster at the Massachusetts Green High-Performance Computing Center (MGHPCC).
This work makes use of the Black Hole Perturbation Toolkit~\cite{BHPToolkit}.

% ----------------------------------------
% --------------- Appendix ---------------
% ----------------------------------------
\appendix
\section{Eccentricity corrections to the RR force} \label{Appendix A}
In this Appendix, we summarize the process employed to compute the 3PN eccentric corrections to the QC RR force~\eqref{QC force} and the expressions for the Schott energy and angular momentum, in the TM limit. A more detailed procedure and the expressions valid for generic mass ratios are presented in Ref.~\cite{Gamboa:2024imd}, based on the ideas and calculations of Refs.~\cite{Bini:2012ji,Khalil:2021txt}. Here, we just outline the basic steps necessary to derive the expressions used in this work.

We start by writing Eqs.~\eqref{eq:balance_equations} and \eqref{eq:dotE_dotJ_system} in the form
  \begin{subequations} 
  \label{eq:forces}
  \begin{align}
    & \dot{r}\mF_{r} = - \Phi_{E} + \dot{\varphi}\Phi_{J}  - \dot{E}_{\text{Schott}} +\dot{\varphi} \dot{J}_{\text{Schott}}, \\
    & \mF_{\varphi}  = - \Phi_{J} - \dot{J}_{\text{Schott}}.
  \end{align}
  \end{subequations}
Given expressions for the fluxes $ \Phi_{E} $ and $ \Phi_{J} $, ansatzes for the Schott terms $ E_{\text{Schott}}  $ and  $ J_{\text{Schott}}  $, and the equations of motion \eqref{Ham_EOM}, we employ Eqs.~\eqref{eq:forces} to determine PN expressions for the radial and azimuthal components of the RR force.

More specifically, we employ the complete 3PN expressions of the EOB fluxes presented in Ref.~\cite{Gamboa:2024imd}. Then, the free unknown coefficients (gauge constants) appearing in the ansatzes for the Schott terms are determined by imposing the regularity of the radial component of the RR force in the QC limit (this means no $ 1/p_r $ terms in $ \mathcal{F}_r $), and by choosing a specific gauge. Following Refs.~\cite{Khalil:2021txt, Gamboa:2024imd} we employ a gauge that satisfies
\begin{subequations}
\label{eq:gauge}
\begin{align}
\mathcal{F}_\varphi
&= - \Phi_\mathrm{J} + \mathcal{O}(p_r^2),
\label{eq:} \\
\frac{\mathcal{F}_r}{p_r}
&= \frac{\mathcal{F}_\varphi}{p_\varphi} + \mathcal{O}(p_r^2).
\end{align}
\end{subequations}
With this choice, the QC limit of the resulting RR force satisfies the gauge used in \texttt{SEOBNRv5HM}, which is given by Eqs.~\eqref{QC force}.

At leading PN order, the components of the RR force, in a generic gauge, are given by 
\begin{subequations}
\label{eq:FrFphiLO}
\begin{align}
\mathcal{F}^\text{LO}_r  &= \frac{8 \nu p_r}{15 \, r^3}  \,  \Big[(-3 \alpha +9 \beta +3) \, p^2 + (9 \alpha -15 \beta +9) \, p_r^2 \nonumber\\
&\quad\qquad 
+ \frac{9 \alpha -9 \beta +17}{r}\Big], \\
 \mathcal{F}^\text{LO}_\varphi  &= \frac{8 \nu p_\varphi }{15 \, r^3}  \, 
\left[9 (\alpha +1) \, p_r^2 -3 (2 + \alpha) \, p^2 + \frac{3 (\alpha -2)}{r}\right],
\end{align}
\end{subequations}
where $ p^2 = p_r^2 + p_\varphi^2/r^2 $, and $\{ \alpha, \beta\}$ are gauge constants, representing the gauge freedom in defining the RR force~\cite{Iyer:1993xi,Iyer:1995rn}.\footnote{
Different values for the gauge constants $ \alpha $ and $ \beta $ have been considered in the literature.
For example:
the choice $\alpha = -1$, $\beta = 0 $ corresponds to the harmonic RR gauge~\cite{Damour:1981ntn};
the values $\alpha = 5/3$, $\beta = 3 $ correspond to the Arnowitt-Deser-Misner (ADM) RR gauge~\cite{Schafer:1982, Arun:2009mc};
and the values $\alpha = 0$, $\beta = 2$ correspond to a choice employed in Ref.~\cite{Bini:2012ji} for a ``minimal decomposition'' of the energy and angular momentum fluxes.
}
In particular, the gauge in Eq.~\eqref{eq:gauge} fixes the value of $ \alpha $ to be $ -16/3 $, but leaves $ \beta $ unspecified. As shown in Ref.~\cite{Gamboa:2024imd}, $\beta$ needs to equal $ -13/2 $ to avoid a 2.5PN modification (relative to the leading order) of the QC orbital phase when transforming between harmonic and EOB coordinates. For the expressions of this work, we use the aforementioned values of $ \alpha $ and $ \beta $, to be consistent with Ref.~\cite{Gamboa:2024imd}. As a consequence of this, the expressions used here do not follow the RR gauge employed in Ref.~\cite{Khalil:2021txt}.

The next step consists in factorizing the RR force in terms of a QC part and an eccentric correction. In this way, given the PN expressions of $ \mathcal{F}_\varphi $ and $ \mathcal{F}_r $, and the QC part given by Eqs.~\eqref{QC force}, we determine the different PN expressions for $ \mathcal{F}_\varphi ^{ \mathrm{ecc}} $ and $ \mathcal{F}_r ^{ \mathrm{ecc}} $ which are specified in Eqs.~\eqref{ecc impl}. The results valid for the TM limit are explicitly shown in Sections \ref{Multiplicative corrections} and \ref{Additive RR force eccentric corrections}. Additionally, the expressions for the time derivatives of the Schott terms in the TM limit are shown in Sec.~\ref{Schott terms}.

In the following subsections, we write the expressions in terms of the variables $\{r, p_{r_*}, v_0\}$, where $r$ is the radial separation, $p_{r_*}$ is the radial momentum conjugate to the tortoise coordinate $ r_* $, and $v_0$ is given by
\begin{equation}
  v_0 = \frac{(1 + \dot{p}_{r_*} r^2)^{1/6}}{\sqrt{r}} \ .
\end{equation}

% ------------ ------------------------------------- ---------------
% ------------ Multplicative corrections expressions ---------------
% ------------ ------------------------------------- ---------------
\begin{widetext}
\subsection{Multiplicative corrections}
\label{Multiplicative corrections}
% ------------ Multiplicative radial component  expressions ---------------
The expressions of the eccentric corrections for the multiplicative implementation (\ref{multiplicative impl}) are given by
\begin{equation}
  \mF_{r}^{\text{ecc,mult}} = \mF_{r}^{\text{0PN}} + \mF_{r}^{\text{1PN}} + \mF_{r}^{{\text{1.5PN,Tail}}} + \mF_{r}^{{\text{1.5PN,SO}}} + \mF_{r}^{{\text{2PN}}} + \mF_{r}^{{\text{2PN,SS}}} + \mF_{r}^{{\text{2.5PN,Tail}}} + \mF_{r}^{{\text{3PN}}} + \mF_{r}^{{\text{3PN,Tail}}} \ ,
\end{equation}
where the expressions of the different PN orders are
 \begin{subequations}
  \begin{equation}
    \mF_{r}^{\text{0PN}} = -\frac{19 p_{r_{*}}^2}{12 r v_0^4}-\frac{55}{24 r^2 v_0^4}+\frac{79
    r v_0^2}{24},
  \end{equation}
  \begin{align}
    \mF_{r}^{\text{1PN}} = & -\frac{19 p_{r_{*}}^4}{36 r^4 v_0^{10}}-\frac{88 p_{r_{*}}^4}{63 r v_0^4}-\frac{131 p_{r_{*}}^2}{72 r^5 v_0^{10}}+\frac{19}{432} p_{r_{*}}^2 r^3 v_0^6-\frac{6851 p_{r_{*}}^2}{224 r^2 v_0^4}-\frac{617 p_{r_{*}}^2 r v_0^2}{1008}-\frac{52459 p_{r_{*}}^2}{12096 r v_0^2}-\frac{55}{36 r^6 v_0^{10}} \nonumber\\ 
    & -\frac{79}{864} r^5 v_0^{12}+\frac{823 r^3 v_0^8}{504}-\frac{52897}{2016 r^3 v_0^4}+\frac{55 r^2 v_0^6}{864}-\frac{151855}{24192 r^2 v_0^2}+\frac{218119 r v_0^4}{24192}+\frac{47219 v_0^2}{2016} \ ,
  \end{align}
  \begin{equation}
    \mF_{r}^{\text{1.5PN,Tail}} = \pi \left[\frac{5  p_{r_{*}}^4}{576 v_0}+\frac{461 p_{r_{*}}^2}{72 r v_0}+\frac{79}{6 r^2 v_0}-\frac{79}{6} r v_0^5 \right] \,,
  \end{equation}
  \begin{equation}
    \mF_{r}^{\text{1.5PN,SO}} = a \left[ \frac{19 p_{r_{*}}^2}{18 r^4 v_0^7}-\frac{19}{144} p_{r_{*}}^2 r^3 v_0^7-\frac{29 p_{r_{*}}^2}{9 r v_0}+\frac{79}{288} r^5
    v_0^{13}+\frac{55}{36 r^5 v_0^7}-\frac{55}{288} r^2 v_0^7-\frac{775}{72 r^2 v_0}+\frac{659}{72} r v_0^5 \right]\,,
  \end{equation}
  \begin{align}
    \mF_{r}^{\text{2PN}} = & -\frac{95 p_{r_{*}}^6}{432 r^7 v_0^{16}}-\frac{211 p_{r_{*}}^6}{3024 r^4 v_0^{10}}+\frac{1013 p_{r_{*}}^6}{864 r v_0^4}-\frac{115 p_{r_{*}}^4}{96 r^8 v_0^{16}}-\frac{1055 p_{r_{*}}^4}{96 r^5 v_0^{10}}-\frac{52459 p_{r_{*}}^4}{72576 r^4 v_0^8}+\frac{13 p_{r_{*}}^4 r^3 v_0^6}{6048}+\frac{605 p_{r_{*}}^4}{84 r^2 v_0^4}+\frac{1129}{672} p_{r_{*}}^4 r v_0^2 \nonumber\\ 
    & -\frac{1576531 p_{r_{*}}^4}{508032 r v_0^2}-\frac{19 p_{r_{*}}^4}{864}-\frac{155 p_{r_{*}}^2}{72 r^9 v_0^{16}}-\frac{19 p_{r_{*}}^2 r^7 v_0^{16}}{15552}-\frac{60605 p_{r_{*}}^2}{2016 r^6 v_0^{10}}+\frac{19}{336} p_{r_{*}}^2 r^5 v_0^{12}-\frac{361691 p_{r_{*}}^2}{145152 r^5 v_0^8}-\frac{86141 p_{r_{*}}^2 r^3 v_0^8}{217728} \nonumber\\ 
    & -\frac{2683325 p_{r_{*}}^2}{12096 r^3 v_0^4}+\frac{9343 p_{r_{*}}^2 r^2 v_0^6}{8064}-\frac{58524917 p_{r_{*}}^2}{677376 r^2 v_0^2}-\frac{311993 p_{r_{*}}^2 r v_0^4}{127008}-\frac{11628649 p_{r_{*}}^2}{451584 r}+\frac{155599 p_{r_{*}}^2 v_0^2}{4032}-\frac{275}{216 r^{10} v_0^{16}} \nonumber\\ 
    & +\frac{79 r^9 v_0^{22}}{31104}+\frac{373 r^7 v_0^{18}}{12096}-\frac{54101}{3024 r^7 v_0^{10}}-\frac{55 r^6 v_0^{16}}{31104}-\frac{151855}{72576 r^6 v_0^8}+\frac{543029 r^5 v_0^{14}}{435456}-\frac{1163 r^4 v_0^{12}}{896} +\frac{3017773 r^3 v_0^{10}}{1016064} \nonumber\\ 
    &-\frac{380027}{2268 r^4 v_0^4}-\frac{158533859}{2032128 r^3 v_0^2}+\frac{9287503 r^2 v_0^8}{435456}-\frac{100860815}{2709504 r^2}+\frac{148133911 r v_0^6}{2709504}+\frac{626551 v_0^2}{4032 r}+\frac{141658627 v_0^4}{2032128} \,,
  \end{align}
  \begin{align}
    \mF_{r}^{\text{2PN,SS}} = &\, a^2 \bigg[-\frac{19 p_{r_{*}}^4}{18 r^5 v_0^{10}}-\frac{55 p_{r_{*}}^2}{36 r^6 v_0^{10}}+\frac{19}{192} p_{r_{*}}^2 r^3v_0^8+\frac{131 p_{r_{*}}^2}{72 r^3 v_0^4}+\frac{19 p_{r_{*}}^2}{6 r}-\frac{79}{384} r^5 v_0^{14}+\frac{145}{96 r^4v_0^4}+\frac{55}{384} r^2 v_0^8 \nonumber\\
    &\quad +\frac{55}{12 r^2}-\frac{79}{12} r v_0^6+\frac{53 v_0^2}{96 r}\bigg] \,,
  \end{align}
  \begin{align}
    \mF_{r}^{{\text{2.5PN,Tail}}} = &\, \pi \bigg[ \frac{5 p_{r_{*}}^6}{6912 r^3 v_0^7}-\frac{413 p_{r_{*}}^6}{34560 v_0}-\frac{5  p_{r_{*}}^4 r^4 v_0^9}{20736}+\frac{1849 p_{r_{*}}^4}{3456 r^4 v_0^7}-\frac{413 p_{r_{*}}^4 r^2 v_0^5}{34560}+\frac{1090127 p_{r_{*}}^4}{241920 r v_0} +\frac{13805  p_{r_{*}}^4v_0}{580608}+\frac{935 p_{r_{*}}^2}{432 r^5 v_0^7} \nonumber\\ 
    &\quad -\frac{689 p_{r_{*}}^2 r^3 v_0^9}{2592}+\frac{261253 p_{r_{*}}^2}{2688r^2 v_0}+\frac{25385 p_{r_{*}}^2 r v_0^5}{6048}+\frac{198055 p_{r_{*}}^2 v_0}{9072 r} +\frac{79}{36 r^6v_0^7}+\frac{79}{144} r^5 v_0^{15} -\frac{1633}{504} r^3v_0^{11} \nonumber\\ 
    &\quad +\frac{201715}{2016 r^3 v_0}-\frac{71}{144} r^2v_0^9+\frac{227293 v_0}{5376 r^2}-\frac{725615 rv_0^7}{16128}-\frac{48563 v_0^5}{504} \bigg]\,,
  \end{align}
  \begin{align}
    \mF_{r}^{{\text{3PN}}} = & -\frac{79 r^{13} v_0^{32}}{1119744}-\frac{971 r^{11} v_0^{28}}{326592}+\frac{19 p_{r_{*}}^2 r^{11} v_0^{26}}{559872}+\frac{55 r^{10} v_0^{26}}{1119744}-\frac{262097 r^9 v_0^{24}}{10450944}-\frac{3487 p_{r_{*}}^2 r^9 v_0^{22}}{1306368}+\frac{141187 r^8 v_0^{22}}{2612736} \nonumber\\ 
    & +\frac{2667851 r^7 v_0^{20}}{12573792}-\frac{278995 p_{r_{*}}^2 r^7 v_0^{18}}{5225472}+\frac{4362881 r^6 v_0^{18}}{10450944}+\frac{313 p_{r_{*}}^4 r^7 v_0^{16}}{326592}-\frac{1315 p_{r_{*}}^2 r^6 v_0^{16}}{32256}+\frac{8201430199 r^5 v_0^{16}}{4389396480} \nonumber\\ 
    & -\frac{290025695 p_{r_{*}}^2 r^5 v_0^{14}}{201180672}-\frac{383131475 r^4 v_0^{14}}{36578304}-\frac{263 p_{r_{*}}^4 r^5 v_0^{12}}{2304}-\frac{18791 p_{r_{*}}^2 r^4 v_0^{12}}{20736}-\frac{318507491 r^3 v_0^{12}}{85349376}+\frac{19 p_{r_{*}}^4 r^4 v_0^{10}}{11664} \nonumber\\ 
    & -\frac{398293579 p_{r_{*}}^2 r^3 v_0^{10}}{2194698240}+\frac{39377759509 r^2 v_0^{10}}{877879296}+\frac{63462257 p_{r_{*}}^4 r^3 v_0^8}{50295168}+\frac{3081350551 p_{r_{*}}^2 r^2 v_0^8}{134120448}+\frac{343837498725233 r v_0^8}{1351934115840} \nonumber\\ 
    & -\frac{2033}{70} r v_0^8 \ln r-\frac{33812}{315} r v_0^8 \ln v_0-\frac{16906 \ln 2}{315} r v_0^8-\frac{71 p_{r_{*}}^6 r^3 v_0^6}{2268}-\frac{47429 p_{r_{*}}^4 r^2 v_0^6}{48384}-\frac{2094514621 p_{r_{*}}^2 r v_0^6}{341397504} \nonumber\\ 
    & +\frac{940303127443 v_0^6}{2048385024}+\frac{996721 p_{r_{*}}^2 v_0^4}{9216}+\frac{24034505 p_{r_{*}}^4 r v_0^4}{4064256}+\frac{15293179 v_0^4}{31752 r}-\frac{68397359 p_{r_{*}}^4 v_0^2}{6531840}-\frac{911609 p_{r_{*}}^6 r v_0^2}{598752} \nonumber\\ 
    & +\frac{1712 p_{r_{*}}^2 v_0^2 \ln r}{35 r}+\frac{2354 v_0^2 \ln r}{45 r^2}+\frac{16264 p_{r_{*}}^2 v_0^2 \ln v_0}{315 r}+\frac{4708 v_0^2 \ln v_0}{63 r^2}-\frac{551068833699073 p_{r_{*}}^2 v_0^2}{3379835289600r} \nonumber\\ 
    & +\frac{2181004862746547 v_0^2}{6759670579200 r^2}+\frac{8132 p_{r_{*}}^2 v_0^2 \ln 2}{315 r}+\frac{2354 v_0^2 \ln 2}{63 r^2}+\frac{373 p_{r_{*}}^6}{24192}-\frac{15435593 p_{r_{*}}^4}{1053696 r}-\frac{13372807411 p_{r_{*}}^2}{25288704 r^2} \nonumber\\ 
    & -\frac{37892126591}{75866112 r^3}+\frac{1305953 p_{r_{*}}^6}{381024 r v_0^2}+\frac{136832399 p_{r_{*}}^4}{4064256 r^2 v_0^2}-\frac{1988331389 p_{r_{*}}^2}{3048192 r^3 v_0^2}-\frac{2744503025}{5225472 r^4 v_0^2}+\frac{856 p_{r_{*}}^4 \ln r}{35 r^3 v_0^4} \nonumber\\ 
    & -\frac{23861 p_{r_{*}}^2 \ln r}{630 r^4 v_0^4}-\frac{24931 \ln r}{630 r^5 v_0^4}-\frac{135077 p_{r_{*}}^8}{149688 r v_0^4}+\frac{32911 p_{r_{*}}^6}{18711 r^2 v_0^4}-\frac{1193113967 p_{r_{*}}^4}{13305600 r^3 v_0^4}-\frac{19358154601 p_{r_{*}}^2}{29937600 r^4 v_0^4} \nonumber\\ 
    & -\frac{112700503}{316800 r^5 v_0^4}+\frac{151855 p_{r_{*}}^6}{6096384 r^4 v_0^8}-\frac{731665 p_{r_{*}}^4}{48384 r^5 v_0^8}-\frac{173012543 p_{r_{*}}^2}{4064256 r^6 v_0^8}-\frac{158804437}{6096384 r^7 v_0^8}+\frac{3721 p_{r_{*}}^8}{9072 r^4 v_0^{10}}+\frac{264151 p_{r_{*}}^6}{24192 r^5 v_0^{10}} \nonumber\\ 
    & -\frac{868811 p_{r_{*}}^4}{12096 r^6 v_0^{10}}-\frac{3011119 p_{r_{*}}^2}{13608 r^7 v_0^{10}}-\frac{804565}{6804 r^8 v_0^{10}}-\frac{52459 p_{r_{*}}^6}{217728 r^7 v_0^{14}}-\frac{63503 p_{r_{*}}^4}{48384 r^8 v_0^{14}}-\frac{85591 p_{r_{*}}^2}{36288 r^9 v_0^{14}}-\frac{151855}{108864 r^{10} v_0^{14}} \nonumber\\ 
    & +\frac{2465 p_{r_{*}}^8}{18144 r^7 v_0^{16}}-\frac{930955 p_{r_{*}}^6}{217728 r^8 v_0^{16}}-\frac{1604525 p_{r_{*}}^4}{72576 r^9 v_0^{16}}-\frac{102055 p_{r_{*}}^2}{3024 r^{10} v_0^{16}}-\frac{840635}{54432 r^{11} v_0^{16}}-\frac{95 p_{r_{*}}^8}{972 r^{10} v_0^{22}}-\frac{1415 p_{r_{*}}^6}{1944 r^{11} v_0^{22}} \nonumber\\ 
    & -\frac{655 p_{r_{*}}^4}{324 r^{12} v_0^{22}}-\frac{1205 p_{r_{*}}^2}{486 r^{13} v_0^{22}}-\frac{275}{243 r^{14} v_0^{22}} \ ,
  \end{align}
  \begin{align}
\mF_{r}^{{\text{3PN,Tail}}} = &\, \frac{5593639 p_{r_{*}}^4}{3150 r^3 v_0^4}-\frac{1044921875 p_{r_{*}}^4 \ln 5}{12096 r^3 v_0^4}-\frac{1061386821 p_{r_{*}}^4 \ln 3}{11200 r^3 v_0^4}+\frac{548384416 p_{r_{*}}^4 \ln 2}{1575 r^3 v_0^4}-\frac{5}{144} \pi ^2 p_{r_{*}}^4 v_0^2 \nonumber\\ 
& -\frac{1870681 p_{r_{*}}^4 v_0^2}{4200}+\frac{208984375 p_{r_{*}}^4 v_0^2 \ln 5}{6048}+\frac{34399323}{800} p_{r_{*}}^4 v_0^2 \ln 3-\frac{698109088 p_{r_{*}}^4 v_0^2 \ln 2}{4725}+\frac{703 \pi ^2 p_{r_{*}}^2}{18 r^4 v_0^4} \nonumber\\ 
& -\frac{75221 \gamma  p_{r_{*}}^2}{630 r^4 v_0^4}+\frac{100258619 p_{r_{*}}^2}{117600 r^4 v_0^4}+\frac{75221 p_{r_{*}}^2 \ln r}{420 r^4 v_0^4}+\frac{1794069 p_{r_{*}}^2 \ln 3}{112 r^4 v_0^4}-\frac{51789391 p_{r_{*}}^2 \ln 2}{1890 r^4 v_0^4}-\frac{103 \pi ^2 p_{r_{*}}^2 v_0^2}{6 r} \nonumber\\ 
& -\frac{8132 \gamma p_{r_{*}}^2 v_0^2}{315 r}-\frac{5498951 p_{r_{*}}^2 v_0^2}{11025 r}-\frac{8132 p_{r_{*}}^2 v_0^2 \ln v_0}{105 r}-\frac{234009 p_{r_{*}}^2 v_0^2 \ln 3}{28r}+\frac{4368596 p_{r_{*}}^2 v_0^2 \ln 2}{315 r}+\frac{413 \pi ^2}{18r^5 v_0^4} \nonumber\\ 
& -\frac{6313 \gamma }{90 r^5 v_0^4}-\frac{109379}{50400 r^5 v_0^4}+\frac{6313 \ln r}{60 r^5 v_0^4}-\frac{234009 \ln 3}{560 r^5 v_0^4}+\frac{176657 \ln 2}{630 r^5 v_0^4}-\frac{1045 \pi^2 v_0^2}{18 r^2} +\frac{10379 \gamma  v_0^2}{630 r^2} \nonumber\\ 
& -\frac{9846541 v_0^2}{117600 r^2}-\frac{33919 v_0^2 \ln r}{420 r^2}-\frac{2354 v_0^2 \ln v_0}{21r^2}+\frac{234009 v_0^2 \ln 3}{560 r^2}-\frac{267821 v_0^2 \ln 2}{630 r^2}+\frac{316}{9} \pi ^2 r v_0^8 \nonumber\\ 
& +\frac{16906}{315} \gamma  rv_0^8+\frac{6857279 r v_0^8}{88200}+\frac{16906}{105} rv_0^8 \ln v_0+\frac{16906 \ln 2}{105} r v_0^8 \ ,
  \end{align}
 \end{subequations}
where $\gamma \approx 0.577$ is the Euler-gamma constant.

% ------------ Multiplicative azimuthal component  expressions ---------------
While for the azimuthal component of the RR force we have
\begin{equation}
    \mF_{\varphi}^{\text{ecc,mult}} = \mF_{\varphi}^{\text{0PN}} + \mF_{\varphi}^{\text{1PN}} + \mF_{\varphi}^{{\text{1.5PN,Tail}}} + \mF_{\varphi}^{{\text{1.5PN,SO}}} + \mF_{\varphi}^{{\text{2PN}}} + \mF_{\varphi}^{{\text{2PN,SS}}} + \mF_{\varphi}^{{\text{2.5PN,Tail}}} + \mF_{\varphi}^{{\text{3PN}}} + \mF_{\varphi}^{{\text{3PN,Tail}}} \,,
  \end{equation}
  with
  \begin{subequations}
  \begin{equation}
  \mF_{\varphi}^{\text{0PN}} = \frac{29 p_{r_{*}}^2}{12 r v_0^4}+\frac{11}{6 r^2 v_0^4}-\frac{5 r v_0^2}{6} \,, 
  \end{equation}
  \begin{align}
  \mF_{\varphi}^{\text{1PN}} = &\, \frac{29 p_{r_{*}}^4}{36 r^4 v_0^{10}} - \frac{58 p_{r_{*}}^4}{63 r v_0^4} + \frac{20 p_{r_{*}}^2}{9 r^5 v_0^{10}} - \frac{29}{432} p_{r_{*}}^2 r^3 v_0^6 - \frac{1977 p_{r_{*}}^2}{224 r^2 v_0^4} - \frac{613}{504} p_{r_{*}}^2 r v_0^2 + \frac{80069 p_{r_{*}}^2}{12096 r v_0^2} + \frac{11}{9 r^6 v_0^{10}} + \frac{5 r^5 v_0^{12}}{216} \nonumber\\
  & + \frac{557 r^3 v_0^8}{1008} - \frac{10687}{2016 r^3 v_0^4} - \frac{11 r^2 v_0^6}{216} + \frac{30371}{6048 r^2 v_0^2} - \frac{13805 r v_0^4}{6048} + \frac{1643 v_0^2}{2016} \,,
  \end{align}
  \begin{equation}
  \mF_{\varphi}^{\text{1.5PN,Tail}} = \pi \left[-\frac{49 p_{r_{*}}^6}{1440 r^2 v_0^7}+\frac{171 p_{r_{*}}^2}{16r^4 v_0^7}-\frac{29 p_{r_{*}}^2}{3 r v_0}+\frac{167}{48r^5 v_0^7}-\frac{109}{16 r^2 v_0}+\frac{10}{3} rv_0^5 \right], 
  \end{equation}
  \begin{equation}
  \mF_{\varphi}^{\text{1.5PN,SO}} = a \left[\frac{29}{144} p_{r_{*}}^2 r^3 v_0^7 -\frac{5 p_{r_{*}}^4}{2 r^3 v_0^7}-\frac{73 p_{r_{*}}^2}{9 r^4 v_0^7}+\frac{89
  p_{r_{*}}^2}{18 r v_0}-\frac{5}{72} r^5 v_0^{13}-\frac{11}{9 r^5 v_0^7}+\frac{11}{72} r^2 v_0^7+\frac{215}{72 r^2
  v_0}-\frac{133}{72} r v_0^5 \right],
  \end{equation}
  \begin{align}
      \mF_{\varphi}^{\text{2PN}} = &\, \frac{145 p_{r_{*}}^6}{432 r^7 v_0^{16}} - \frac{2755 p_{r_{*}}^6}{3024 r^4 v_0^{10}} + \frac{37 p_{r_{*}}^6}{108 r v_0^4} + \frac{115 p_{r_{*}}^4}{72 r^8 v_0^{16}} - \frac{6515 p_{r_{*}}^4}{2016 r^5 v_0^{10}} + \frac{80069 p_{r_{*}}^4}{72576 r^4 v_0^8} + \frac{493 p_{r_{*}}^4 r^3 v_0^6}{6048} + \frac{101 p_{r_{*}}^4}{504 r^2 v_0^4} \nonumber\\ 
      & + \frac{505}{252} p_{r_{*}}^4 r v_0^2 - \frac{1841587 p_{r_{*}}^4}{508032 r v_0^2} + \frac{29 p_{r_{*}}^4}{864} + \frac{85 p_{r_{*}}^2}{36 r^9 v_0^{16}} + \frac{29 p_{r_{*}}^2 r^7 v_0^{16}}{15552} - \frac{12755 p_{r_{*}}^2}{2016 r^6 v_0^{10}} + \frac{71 p_{r_{*}}^2 r^5 v_0^{12}}{1008} + \frac{13805 p_{r_{*}}^2}{4536 r^5 v_0^8} \nonumber\\
      &- \frac{261869 p_{r_{*}}^2 r^3 v_0^8}{217728} - \frac{797813 p_{r_{*}}^2}{12096 r^3 v_0^4} - \frac{1355 p_{r_{*}}^2 r^2 v_0^6}{8064} - \frac{14442791 p_{r_{*}}^2}{677376 r^2 v_0^2} - \frac{1029853 p_{r_{*}}^2 r v_0^4}{254016} + \frac{17738551 p_{r_{*}}^2}{451584 r} \nonumber\\
      & + \frac{5311 p_{r_{*}}^2 v_0^2}{576} + \frac{55}{54 r^{10} v_0^{16}} - \frac{5 r^9 v_0^{22}}{7776} - \frac{419 r^7 v_0^{18}}{12096} - \frac{9119}{3024 r^7 v_0^{10}} + \frac{11 r^6 v_0^{16}}{7776} + \frac{30371}{18144 r^6 v_0^8} + \frac{23543 r^5 v_0^{14}}{108864} \nonumber\\
      & + \frac{1373 r^4 v_0^{12}}{8064} - \frac{235303}{18144 r^4 v_0^4} + \frac{1924417 r^3 v_0^{10}}{1016064} - \frac{22858319}{2032128 r^3 v_0^2} - \frac{13733 r^2 v_0^8}{27216} + \frac{20172163}{677376 r^2}  - \frac{9304097 r v_0^6}{677376} \nonumber\\ 
      & + \frac{26207 v_0^2}{4032 r} + \frac{516307 v_0^4}{2032128} \, ,    
  \end{align}
  \begin{align}
      \mF_{\varphi}^{\text{2PN,SS}} = &\, a^2 \bigg[\frac{29 p_{r_{*}}^4}{18 r^5 v_0^{10}}+\frac{11 p_{r_{*}}^2}{9 r^6 v_0^{10}}-\frac{29}{192} p_{r_{*}}^2 r^3v_0^8+\frac{755 p_{r_{*}}^2}{72 r^3 v_0^4}-\frac{29 p_{r_{*}}^2}{6 r}+\frac{5}{96} r^5 v_0^{14} +\frac{409}{96 r^4v_0^4}-\frac{11}{96} r^2 v_0^8 \\
      &\quad -\frac{11}{3 r^2}+\frac{5}{3} r v_0^6-\frac{211 v_0^2}{96 r} \bigg],
  \end{align}
  \begin{align}
      \mF_{\varphi}^{\text{2.5PN,Tail}} = &\, \pi \bigg[\frac{49 p_{r_{*}}^8}{3456 r^2 v_0^7} -\frac{343 p_{r_{*}}^8}{17280 r^5 v_0^{13}} - \frac{343 p_{r_{*}}^6}{8640 r^6 v_0^{13}} - \frac{15361 p_{r_{*}}^6}{30240 r^3 v_0^7} - \frac{19327 p_{r_{*}}^6}{207360 r^2 v_0^5} + \frac{49 p_{r_{*}}^6 r^2 v_0^3}{51840} + \frac{49 p_{r_{*}}^6}{3456 v_0} \nonumber\\ 
      &\quad + \frac{399 p_{r_{*}}^4}{64 r^7 v_0^{13}} + \frac{60827 p_{r_{*}}^4}{5376 r^4 v_0^7} + \frac{1537 p_{r_{*}}^4}{252 r v_0} + \frac{8351 p_{r_{*}}^2}{576 r^8 v_0^{13}} + \frac{184447 p_{r_{*}}^2}{4032 r^5 v_0^7}  + \frac{52459 p_{r_{*}}^2}{1792 r^4 v_0^5} + \frac{29}{72} p_{r_{*}}^2 r^3 v_0^9  - \frac{19}{64} p_{r_{*}}^2 v_0^3 \nonumber\\ 
      &\quad + \frac{12683 p_{r_{*}}^2}{504 r^2 v_0} + \frac{1625}{252} p_{r_{*}}^2 r v_0^5 - \frac{266365 p_{r_{*}}^2 v_0}{8064 r} + \frac{1169}{288 r^9 v_0^{13}} + \frac{1615}{336 r^6 v_0^7} - \frac{5}{36} \pi r^5 v_0^{15} + \frac{461087}{48384 r^5 v_0^5} \nonumber\\ 
      &\quad  - \frac{767}{252} r^3 v_0^{11} - \frac{16217}{4032 r^3 v_0} + \frac{503 r^2 v_0^9}{1728} - \frac{1143395 v_0}{48384 r^2} + \frac{45925 r v_0^7}{4032} - \frac{167 v_0^3}{1728 r} + \frac{1151 v_0^5}{1344} \bigg],
  \end{align}
\begin{align}
    \mF_{\varphi}^{\text{3PN}} = &\, \frac{5 r^{13} v_0^{32}}{279936} + \frac{1957 r^{11} v_0^{28}}{1306368} - \frac{29 p_{r_{*}}^2 r^{11} v_0^{26}}{559872} - \frac{11 r^{10} v_0^{26}}{279936} + \frac{36403 r^9 v_0^{24}}{2612736} - \frac{1943 p_{r_{*}}^2 r^9 v_0^{22}}{653184} - \frac{26357 r^8 v_0^{22}}{2612736} \nonumber\\
    & - \frac{45275695 r^7 v_0^{20}}{201180672} + \frac{17195 p_{r_{*}}^2 r^7 v_0^{18}}{746496} - \frac{707257 r^6 v_0^{18}}{2612736} - \frac{1247 p_{r_{*}}^4 r^7 v_0^{16}}{326592} + \frac{4687 p_{r_{*}}^2 r^6 v_0^{16}}{290304} + \frac{17132501 r^5 v_0^{16}}{31352832} \nonumber\\
    & + \frac{74723225 p_{r_{*}}^2 r^5 v_0^{14}}{100590336} + \frac{915435407 r^4 v_0^{14}}{402361344} - \frac{35}{216} p_{r_{*}}^4 r^5 v_0^{12} + \frac{23329 p_{r_{*}}^2 r^4 v_0^{12}}{145152} + \frac{4837213015 r^3 v_0^{12}}{341397504} - \frac{29 p_{r_{*}}^4 r^4 v_0^{10}}{11664} \nonumber\\
    & - \frac{8345573611 p_{r_{*}}^2 r^3 v_0^{10}}{2194698240} + \frac{171479 r^2 v_0^{10}}{22394880} + \frac{86644739 p_{r_{*}}^4 r^3 v_0^8}{50295168} - \frac{1129077875 p_{r_{*}}^2 r^2 v_0^8}{134120448} + \frac{118884442364929 r v_0^8}{1689917644800} \nonumber\\
    &  - \frac{214}{35} r v_0^8 \ln r + \frac{1712}{63} r v_0^8 \ln v_0 + \frac{856 \ln 2}{63} r v_0^8 - \frac{5959 p_{r_{*}}^6 r^3 v_0^6}{72576} + \frac{13715 p_{r_{*}}^4 r^2 v_0^6}{48384} - \frac{4589832917 p_{r_{*}}^2 r v_0^6}{170698752}  \nonumber\\
    & - \frac{21850038725 v_0^6}{2048385024} + \frac{32151845 p_{r_{*}}^2 v_0^4}{1354752} + \frac{295427 p_{r_{*}}^4 r v_0^4}{42336} + \frac{20710261 v_0^4}{1354752 r} - \frac{193847585 p_{r_{*}}^4 v_0^2}{14370048} - \frac{3322987 p_{r_{*}}^6 r v_0^2}{1197504} \nonumber\\
    & + \frac{2889 p_{r_{*}}^2 v_0^2 \ln r}{70 r} - \frac{535 v_0^2 \ln r}{126 r^2} - \frac{24824 p_{r_{*}}^2 v_0^2 \ln v_0}{315 r} - \frac{18832 v_0^2 \ln v_0}{315 r^2} - \frac{252830224115069 p_{r_{*}}^2 v_0^2}{675967057920 r}  - \frac{9416 v_0^2 \ln 2}{315 r^2} \nonumber\\
    & + \frac{41138929379561 v_0^2}{1689917644800 r^2} - \frac{12412 p_{r_{*}}^2 v_0^2 \ln 2}{315 r} - \frac{1595 p_{r_{*}}^6}{24192} - \frac{29482781 p_{r_{*}}^4}{1053696 r} - \frac{2756994649 p_{r_{*}}^2}{25288704 r^2} - \frac{3565629869}{75866112 r^3}  \nonumber\\
    & + \frac{12206381 p_{r_{*}}^6}{6096384 r v_0^2}  + \frac{527351 p_{r_{*}}^4}{193536 r^2 v_0^2} - \frac{560474717 p_{r_{*}}^2}{3048192 r^3 v_0^2} - \frac{1390812335}{36578304 r^4 v_0^2} - \frac{214 p_{r_{*}}^4 \ln r}{35 r^3 v_0^4} - \frac{7276 p_{r_{*}}^2 \ln r}{315 r^4 v_0^4} - \frac{107 \ln r}{18 r^5 v_0^4}\nonumber\\
    & + \frac{223 p_{r_{*}}^8}{37422 r v_0^4} + \frac{23518547 p_{r_{*}}^6}{5987520 r^2 v_0^4} + \frac{471959591 p_{r_{*}}^4}{4435200 r^3 v_0^4} - \frac{92174527 p_{r_{*}}^2}{1871100 r^4 v_0^4} - \frac{1837807}{190080 r^5 v_0^4} - \frac{8727521 p_{r_{*}}^6}{6096384 r^4 v_0^8} - \frac{17678683 p_{r_{*}}^4}{4064256 r^5 v_0^8} \nonumber\\
    & - \frac{29650379 p_{r_{*}}^2}{4064256 r^6 v_0^8} - \frac{19302151}{6096384 r^7 v_0^8} + \frac{15383 p_{r_{*}}^8}{18144 r^4 v_0^{10}} + \frac{18511 p_{r_{*}}^6}{24192 r^5 v_0^{10}} - \frac{145343 p_{r_{*}}^4}{6048 r^6 v_0^{10}} - \frac{2801305 p_{r_{*}}^2}{54432 r^7 v_0^{10}} - \frac{133409}{13608 r^8 v_0^{10}} \nonumber\\
    & + \frac{80069 p_{r_{*}}^6}{217728 r^7 v_0^{14}} + \frac{63503 p_{r_{*}}^4}{36288 r^8 v_0^{14}} + \frac{46937 p_{r_{*}}^2}{18144 r^9 v_0^{14}} + \frac{30371}{27216 r^{10} v_0^{14}} - \frac{11455 p_{r_{*}}^8}{18144 r^7 v_0^{16}} - \frac{537295 p_{r_{*}}^6}{217728 r^8 v_0^{16}} - \frac{324335 p_{r_{*}}^4}{72576 r^9 v_0^{16}}  \nonumber\\
    & - \frac{970 p_{r_{*}}^2}{189 r^{10} v_0^{16}} - \frac{110465}{54432 r^{11} v_0^{16}} + \frac{145 p_{r_{*}}^8}{972 r^{10} v_0^{22}} + \frac{245 p_{r_{*}}^6}{243 r^{11} v_0^{22}} + \frac{200 p_{r_{*}}^4}{81 r^{12} v_0^{22}} + \frac{620 p_{r_{*}}^2}{243 r^{13} v_0^{22}}  + \frac{220}{243 r^{14} v_0^{22}}  \ ,
\end{align}
\begin{align}
    \mF_{\varphi}^{{\text{3PN, Tail}}} = &\, \frac{49 \pi ^2 p_{r_{*}}^6}{360 r^2 v_0^4} + \frac{1870681 p_{r_{*}}^6}{4200 r^2 v_0^4} - \frac{208984375 p_{r_{*}}^6 \ln 5}{6048 r^2 v_0^4} - \frac{34399323 p_{r_{*}}^6 \ln 3}{800 r^2 v_0^4} + \frac{698109088 p_{r_{*}}^6 \ln 2}{4725 r^2 v_0^4} + \frac{193777 p_{r_{*}}^4}{420 r^3 v_0^4} \nonumber\\
    & + \frac{234009 p_{r_{*}}^4 \ln 3}{28 r^3 v_0^4} - \frac{4392992 p_{r_{*}}^4 \ln 2}{315 r^3 v_0^4} - \frac{1681 \pi ^2 p_{r_{*}}^2}{36 r^4 v_0^4}  + \frac{7597 \gamma  p_{r_{*}}^2}{630 r^4 v_0^4} + \frac{29951111 p_{r_{*}}^2}{352800 r^4 v_0^4} - \frac{7597 p_{r_{*}}^2 \ln r}{420 r^4 v_0^4} \nonumber\\
    & + \frac{78003 p_{r_{*}}^2 \ln 3}{560 r^4 v_0^4} - \frac{69443 p_{r_{*}}^2 \ln 2}{630 r^4 v_0^4}  + \frac{232 \pi ^2 p_{r_{*}}^2 v_0^2}{9 r} + \frac{12412 \gamma  p_{r_{*}}^2 v_0^2}{315 r} + \frac{2517229 p_{r_{*}}^2 v_0^2}{44100 r} + \frac{12412 p_{r_{*}}^2 v_0^2 \ln v_0}{105 r} \nonumber\\
    &  + \frac{12412 p_{r_{*}}^2 v_0^2 \ln 2}{105 r} - \frac{503 \pi ^2}{36 r^5 v_0^4} + \frac{107 \gamma }{630 r^5 v_0^4} - \frac{8511719}{352800 r^5 v_0^4} - \frac{107 \ln r}{420 r^5 v_0^4} + \frac{78003 \ln 3}{560 r^5 v_0^4} - \frac{17441 \ln 2}{126 r^5 v_0^4}+ \frac{823 \pi ^2 v_0^2}{36 r^2} \nonumber\\
    &  + \frac{8453 \gamma  v_0^2}{630 r^2} + \frac{4193213 v_0^2}{117600 r^2} + \frac{10379 v_0^2 \ln r}{420 r^2} + \frac{9416 v_0^2 \ln v_0}{105 r^2} - \frac{78003 v_0^2 \ln 3}{560 r^2}+ \frac{123157 v_0^2 \ln 2}{630 r^2} - \frac{80}{9} \pi ^2 r v_0^8  \nonumber\\
    & - \frac{856}{63} \gamma  r v_0^8 - \frac{86801 r v_0^8}{4410} - \frac{856}{21} r v_0^8 \ln v_0 - \frac{856 \ln 2}{21} r v_0^8 \ .   
\end{align}
\end{subequations}

% ------------ ------------------------------------- ---------------
% ------------ Additive corrections expressions ---------------
% ------------ ------------------------------------- ---------------

\subsection{Additive RR force eccentric corrections}
\label{Additive RR force eccentric corrections}
% ------------ Additive radial component  expressions ---------------
Here we provide the additive implementation (\ref{additive impl}) expressions. For the radial component we have
  \begin{equation}
      \mF_{r}^{\text{ecc,add}} = \mF_{r}^{\text{0PN}} + \mF_{r}^{\text{1PN}} + \mF_{r}^{{\text{1.5PN,Tail}}} + \mF_{r}^{{\text{1.5PN,SO}}} + \mF_{r}^{{\text{2PN}}} + \mF_{r}^{{\text{2PN,SS}}} + \mF_{r}^{{\text{2.5PN,Tail}}} + \mF_{r}^{{\text{3PN}}} + \mF_{r}^{{\text{3PN,Tail}}} \ ,
  \end{equation}
where the different terms are
  \begin{subequations}
  \begin{equation}
  \mF_{r}^{\text{0PN}} = \frac{152 p_{r_{*}}^3}{15 r^3}+\frac{44 p_{r_{*}}}{3 r^4}+\frac{32 p_{r_{*}} v_0^4}{5 r^2}-\frac{316 p_{r_{*}} v_0^6}{15 r} \,,
  \end{equation}
  \begin{align}
      \mF_{r}^{\text{1PN}} = &\, \frac{372 p_{r_{*}}^5}{35 r^3} - \frac{32 p_{r_{*}}^3}{15 r^5 v_0^2} + \frac{21547 p_{r_{*}}^3}{105 r^4} + \frac{16 p_{r_{*}}^3 v_0^4}{15 r^2} + \frac{44 p_{r_{*}}^3 v_0^6}{21 r} - \frac{64 p_{r_{*}}}{15 r^6 v_0^2} + \frac{6369 p_{r_{*}}}{35 r^5} + \frac{8}{45} p_{r_{*}} r^2 v_0^{14} \nonumber\\ 
      & - \frac{51971 p_{r_{*}} v_0^6}{315 r^2} - \frac{1466}{105} p_{r_{*}} r v_0^{12} + \frac{16 p_{r_{*}} v_0^{10}}{15} \,,
  \end{align}
  \begin{equation}
  \mF_{r}^{\text{1.5PN,Tail}} = \pi \left[ -\frac{p_{r_{*}}^5 v_0^3}{18 r^2}-\frac{4 p_{r_{*}}^3 v_0^3}{9 r^3}-\frac{128 p_{r_{*}} v_0^3}{5 r^4}+\frac{128 p_{r_{*}} v_0^7}{5 r^2} \right],
\end{equation}
\begin{equation}
  \mF_{r}^{\text{1.5PN,SO}} = a \left[-\frac{32 p_{r_{*}}^3 v_0^3}{5 r^3}+\frac{64 p_{r_{*}} v_0}{15 r^5}+\frac{236 p_{r_{*}} v_0^3}{15 r^4}-\frac{8}{15} p_{r_{*}} r^2 v_0^{15}-\frac{256 p_{r_{*}} v_0^7}{15 r^2}-\frac{12 p_{r_{*}} v_0^9}{5 r} \right],
\end{equation}
\begin{align}
  \mF_{r}^{\text{2PN}} = & -\frac{2116 p_{r_{*}}^7}{315 r^3}-\frac{8 p_{r_{*}}^5}{45 r^8 v_0^8}+\frac{56 p_{r_{*}}^5}{45 r^5 v_0^2}-\frac{1468 p_{r_{*}}^5}{105 r^4}-\frac{4 p_{r_{*}}^5 v_0^4}{9 r^2} -\frac{1619 p_{r_{*}}^5 v_0^6}{210 r}-\frac{32 p_{r_{*}}^3}{45 r^9 v_0^8}-\frac{128 p_{r_{*}}^3}{45 r^6 v_0^2}+\frac{1440472 p_{r_{*}}^3}{945 r^5} \nonumber\\ 
  & +\frac{152 p_{r_{*}}^3 v_0^4}{45 r^3} -\frac{16}{135} p_{r_{*}}^3 r^2 v_0^{14}-\frac{151853 p_{r_{*}}^3 v_0^6}{630 r^2}+\frac{137}{35} p_{r_{*}}^3 r v_0^{12}-\frac{28 p_{r_{*}}^3 v_0^8}{135 r}-\frac{16 p_{r_{*}}^3 v_0^{10}}{45} -\frac{32 p_{r_{*}}}{45 r^{10} v_0^8}-\frac{32 p_{r_{*}}}{15 r^7 v_0^2} \nonumber\\ 
  & +\frac{3403531 p_{r_{*}}}{2835 r^6}-\frac{16}{135} p_{r_{*}} r^4 v_0^{20}+\frac{8 p_{r_{*}} v_0^4}{9 r^4}-\frac{7829}{630} p_{r_{*}} r^3 v_0^{18} -\frac{908602 p_{r_{*}} v_0^6}{945 r^3}-\frac{16}{105} p_{r_{*}} r^2 v_0^{16} -\frac{159268 p_{r_{*}} v_0^8}{2835 r^2} \nonumber\\ 
  & +\frac{52}{45} p_{r_{*}} r v_0^{14}+\frac{64 p_{r_{*}} v_0^{10}}{15 r}-\frac{109397 p_{r_{*}} v_0^{12}}{630} \ , 
\end{align}
\begin{equation}
  \mF_{r}^{\text{2PN,SS}} = a^2 \left[-\frac{64 p_{r_{*}}^3}{15 r^6 v_0^2}-\frac{116 p_{r_{*}}^3}{15 r^5}-\frac{73 p_{r_{*}}}{3 r^6}-\frac{32 p_{r_{*}} v_0^4}{5r^4}+\frac{263 p_{r_{*}} v_0^6}{15 r^3}+\frac{2}{5} p_{r_{*}} r^2 v_0^{16}+\frac{64 p_{r_{*}} v_0^8}{5 r^2} \right],
\end{equation}
\begin{align}
\mF_{r}^{\text{2.5PN,Tail}} = &\, \pi \bigg[ \frac{p_{r_{*}}^7}{72 r^5 v_0^3} + \frac{121 p_{r_{*}}^7 v_0^3}{1800 r^2} + \frac{5 p_{r_{*}}^5}{36 r^6 v_0^3} - \frac{41429 p_{r_{*}}^5 v_0^3}{12600 r^3} + \frac{121 p_{r_{*}}^5 v_0^9}{1800} + \frac{298 p_{r_{*}}^3}{45 r^7 v_0^3} - \frac{224 p_{r_{*}}^3 v_0}{15 r^5}+ \frac{81259 p_{r_{*}}^3 v_0^3}{420 r^4} - \frac{p_{r_{*}}^3 v_0^9}{3 r} \nonumber\\
&\quad  - \frac{32 p_{r_{*}}^3 v_0^7}{15 r^2} + \frac{64 p_{r_{*}}}{5 r^8 v_0^3} - \frac{448 p_{r_{*}} v_0}{15 r^6} + \frac{2467 p_{r_{*}} v_0^3}{21 r^5} + \frac{96 p_{r_{*}} v_0^7}{5 r^3} + \frac{16}{45} p_{r_{*}} r^2 v_0^{17} - \frac{37117 p_{r_{*}} v_0^9}{315 r^2} - \frac{32}{15} p_{r_{*}} v_0^{13} \bigg],
\end{align}
\begin{align}
  \mF_{r}^{\text{3PN}} = &\, \frac{15017 p_{r_{*}}^9}{3465 r^3} - \frac{16 p_{r_{*}}^7}{405 r^{11} v_0^{14}} + \frac{32 p_{r_{*}}^7}{135 r^8 v_0^8} - \frac{124 p_{r_{*}}^7}{135 r^5 v_0^2} - \frac{666419 p_{r_{*}}^7}{27720 r^4} + \frac{22 p_{r_{*}}^7 v_0^4}{81 r^2} + \frac{37519 p_{r_{*}}^7 v_0^6}{6930 r} - \frac{32 p_{r_{*}}^5}{135 r^{12} v_0^{14}} + \frac{8 p_{r_{*}}^5}{135 r^9 v_0^8} \nonumber\\
  & + \frac{52 p_{r_{*}}^5}{15 r^6 v_0^2} + \frac{503253523 p_{r_{*}}^5}{519750 r^5} - \frac{27392 p_{r_{*}}^5 \ln r}{175 r^5} + \frac{28 p_{r_{*}}^5 v_0^2}{405 r^4} - \frac{70 p_{r_{*}}^5 v_0^4}{27 r^3} + \frac{8}{81} p_{r_{*}}^5 r^2 v_0^{14} + \frac{19473 p_{r_{*}}^5 v_0^6}{700 r^2} - \frac{22301 p_{r_{*}}^5 r v_0^{12}}{2310} \nonumber\\
  & + \frac{119 p_{r_{*}}^5 v_0^8}{405 r} + \frac{32 p_{r_{*}}^5 v_0^{10}}{135} - \frac{64 p_{r_{*}}^3}{135 r^{13} v_0^{14}} - \frac{32 p_{r_{*}}^3}{27 r^{10} v_0^8} - \frac{904 p_{r_{*}}^3}{135 r^7 v_0^2} + \frac{3971393821 p_{r_{*}}^3}{779625 r^6} + \frac{381776 p_{r_{*}}^3 \ln r}{1575 r^6} + \frac{318536 p_{r_{*}}^3 v_0^2}{8505 r^5} \nonumber\\
  & + \frac{56}{405} p_{r_{*}}^3 r^4 v_0^{20} + \frac{1412 p_{r_{*}}^3 v_0^4}{135 r^4} + \frac{29455 p_{r_{*}}^3 r^3 v_0^{18}}{2772} + \frac{1662031157 p_{r_{*}}^3 v_0^6}{1039500 r^3} - \frac{54784 p_{r_{*}}^3 v_0^6 \ln r}{175 r^3}+ \frac{68168 p_{r_{*}}^3 v_0^8}{8505 r^2}  \nonumber\\
  &+ \frac{8}{63} p_{r_{*}}^3 r^2 v_0^{16}  - \frac{496 p_{r_{*}}^3 v_0^{10}}{315 r}  - \frac{244}{405} p_{r_{*}}^3 r v_0^{14} - \frac{4446151 p_{r_{*}}^3 v_0^{12}}{27720} - \frac{128 p_{r_{*}}}{405 r^{14} v_0^{14}} - \frac{32 p_{r_{*}}}{45 r^{11} v_0^8}  - \frac{448 p_{r_{*}}}{135 r^8 v_0^2} + \frac{16}{405} p_{r_{*}} r^6 v_0^{26}  \nonumber\\
  & + \frac{2392124741 p_{r_{*}}}{779625 r^7} + \frac{398896 p_{r_{*}} \ln r}{1575 r^7} + \frac{126944 p_{r_{*}} v_0^2}{1701 r^6} - \frac{1871}{990} p_{r_{*}} r^5 v_0^{24} + \frac{458 p_{r_{*}} r^4 v_0^{22}}{2835} - \frac{1686962227 p_{r_{*}} v_0^6}{1559250 r^4} \nonumber\\
  &  - \frac{75328 p_{r_{*}} v_0^6 \ln r}{225 r^4} - \frac{17}{15} p_{r_{*}} r^3 v_0^{20} - \frac{166324 p_{r_{*}} v_0^8}{2835 r^3} - \frac{41657 p_{r_{*}} r^2 v_0^{18}}{8100} + \frac{560238598 p_{r_{*}} v_0^{10}}{779625 r^2} + \frac{109568 p_{r_{*}} v_0^{10} \ln v_0}{525 r^2} \nonumber\\
  & + \frac{54784 p_{r_{*}} v_0^{10} \ln 2}{525 r^2} - \frac{38}{45} p_{r_{*}} r v_0^{16} - \frac{320493581 p_{r_{*}} v_0^{12}}{115500 r} + \frac{32528 p_{r_{*}} v_0^{12} \ln r}{175 r} + \frac{25276 p_{r_{*}} v_0^{14}}{1701} \ ,
\end{align}
\begin{align}
  \mF_{r}^{\text{3PN, Tail}} = & -\frac{89498224 p_{r_{*}}^5}{7875 r^5} + \frac{208984375 p_{r_{*}}^5 \ln 5}{378 r^5} + \frac{1061386821 p_{r_{*}}^5 \ln 3}{1750 r^5} - \frac{17548301312 p_{r_{*}}^5 \ln 2}{7875 r^5} + \frac{7482724 p_{r_{*}}^5 v_0^6}{2625 r^2} \nonumber\\
  & - \frac{41796875 p_{r_{*}}^5 v_0^6 \ln 5}{189 r^2} - \frac{34399323 p_{r_{*}}^5 v_0^6 \ln 3}{125 r^2} + \frac{22339490816 p_{r_{*}}^5 v_0^6 \ln 2}{23625 r^2} - \frac{11248 \pi ^2 p_{r_{*}}^3}{45 r^6} + \frac{1203536 \gamma  p_{r_{*}}^3}{1575 r^6} \nonumber\\
  & - \frac{100258619 p_{r_{*}}^3}{18375 r^6} - \frac{601768 p_{r_{*}}^3 \ln r}{525 r^6} - \frac{3588138 p_{r_{*}}^3 \ln 3}{35 r^6} + \frac{828630256 p_{r_{*}}^3 \ln 2}{4725 r^6} + \frac{1550216 p_{r_{*}}^3 v_0^6}{525 r^3}  - \frac{6608 \pi ^2 p_{r_{*}}}{45 r^7} \nonumber\\
  &  + \frac{1872072 p_{r_{*}}^3 v_0^6 \ln 3}{35 r^3} - \frac{140575744 p_{r_{*}}^3 v_0^6 \ln 2}{1575 r^3} + \frac{101008 \gamma  p_{r_{*}}}{225 r^7} + \frac{109379 p_{r_{*}}}{7875 r^7} - \frac{50504 p_{r_{*}} \ln r}{75 r^7} + \frac{468018 p_{r_{*}} \ln 3}{175 r^7} \nonumber\\
  & - \frac{2826512 p_{r_{*}} \ln 2}{1575 r^7} + \frac{5072 \pi ^2 p_{r_{*}} v_0^6}{45 r^4} - \frac{542704 \gamma  p_{r_{*}} v_0^6}{1575 r^4} + \frac{10443403 p_{r_{*}} v_0^6}{55125 r^4} + \frac{271352 p_{r_{*}} v_0^6 \ln r}{525 r^4}  + \frac{512 \pi ^2 p_{r_{*}} v_0^{10}}{15 r^2} \nonumber\\
  & - \frac{468018 p_{r_{*}} v_0^6 \ln 3}{175 r^4} + \frac{3155216 p_{r_{*}} v_0^6 \ln 2}{1575 r^4} - \frac{54784 \gamma  p_{r_{*}} v_0^{10}}{525 r^2}  - \frac{2777632 p_{r_{*}} v_0^{10}}{18375 r^2} - \frac{54784 p_{r_{*}} v_0^{10} \ln v_0}{175 r^2} \nonumber\\
  & - \frac{54784 p_{r_{*}} v_0^{10} \ln 2}{175 r^2} \ .
\end{align}
\end{subequations}

% ------------ Additive azimuthal component  expressions ---------------
The azimuthal component is
\begin{equation}
  \mF_{\varphi}^{\text{ecc,add}} = \mF_{\varphi}^{\text{0PN}} + \mF_{\varphi}^{\text{1PN}} + \mF_{\varphi}^{{\text{1.5PN,Tail}}} + \mF_{\varphi}^{{\text{1.5PN,SO}}} + \mF_{\varphi}^{{\text{2PN}}} + \mF_{\varphi}^{{\text{2PN,SS}}} + \mF_{\varphi}^{{\text{2.5PN,Tail}}} + \mF_{\varphi}^{{\text{3PN}}} + \mF_{\varphi}^{{\text{3PN,Tail}}} \ ,
\end{equation}
while all the different PN orders are given by
\begin{subequations}
\begin{equation}
\mF_{\varphi}^{\text{0PN}} = -\frac{232 p_{r_{*}}^2 v_0^3}{15 r}-\frac{176 v_0^3}{15 r^2}+\frac{16 r v_0^9}{3}+\frac{32 v_0^7}{5} \ , \\
\end{equation}
\begin{align}
\mF_{\varphi}^{\text{1PN}} = &\, \frac{58 p_{r_{*}}^4}{15 r^4 v_0^3} - \frac{58 p_{r_{*}}^4 v_0^3}{105 r} + \frac{32 p_{r_{*}}^2}{3 r^5 v_0^3} - \frac{56 p_{r_{*}}^2 v_0}{15 r^3} + \frac{5497 p_{r_{*}}^2 v_0^3}{105 r^2} + \frac{374}{105} p_{r_{*}}^2 r v_0^9 + \frac{8 p_{r_{*}}^2 v_0^7}{3} + \frac{88}{15 r^6 v_0^3} \nonumber\\
& + \frac{8 r^4 v_0^{17}}{45} - \frac{112 v_0}{15 r^4} - \frac{46 r^3 v_0^{15}}{35} + \frac{3217 v_0^3}{105 r^3} + \frac{8 r^2 v_0^{13}}{3} - \frac{8 v_0^7}{5 r} - \frac{1825 v_0^9}{63} \ , 
\end{align}
\begin{equation}
\mF_{\varphi}^{\text{1.5PN,Tail}} = \pi \left[\frac{49 p_{r_{*}}^6}{225 r^2}-\frac{342 p_{r_{*}}^2}{5 r^4}-\frac{334
}{15 r^5}-\frac{10 v_0^6}{3 r^2}+\frac{128 
v_0^{10}}{5} \right],
\end{equation} 
\begin{equation}
\mF_{\varphi}^{\text{1.5PN,SO}} = a \left[ \frac{16 p_{r_{*}}^4}{r^3}+\frac{88 p_{r_{*}}^2}{r^4}+\frac{48 p_{r_{*}}^2 v_0^6}{5 r}+\frac{176}{5 r^5}-\frac{8}{15} r^4v_0^{18}-\frac{224 v_0^4}{15 r^3}-\frac{4 v_0^6}{15 r^2}-\frac{12}{5} r v_0^{12}-\frac{256 v_0^{10}}{15} \right],
\end{equation} 
\begin{align}
\mF_{\varphi}^{\text{2PN}} = &\, \frac{29 p_{r_{*}}^6}{60 r^7 v_0^9} - \frac{58 p_{r_{*}}^6}{21 r^4 v_0^3} + \frac{2699 p_{r_{*}}^6 v_0^3}{1260 r} + \frac{23 p_{r_{*}}^4}{10 r^8 v_0^9} + \frac{7 p_{r_{*}}^4}{45 r^6 v_0^5} - \frac{4891 p_{r_{*}}^4}{420 r^5 v_0^3} + \frac{56 p_{r_{*}}^4 v_0}{45 r^3} + \frac{453 p_{r_{*}}^4 v_0^3}{140 r^2} - \frac{152}{21} p_{r_{*}}^4 r v_0^9 \nonumber\\
& - \frac{7 p_{r_{*}}^4 v_0^7}{9} + \frac{17 p_{r_{*}}^2}{5 r^9 v_0^9} + \frac{28 p_{r_{*}}^2}{45 r^7 v_0^5} - \frac{813 p_{r_{*}}^2}{28 r^6 v_0^3} - \frac{2}{27} p_{r_{*}}^2 r^4 v_0^{17} - \frac{266 p_{r_{*}}^2 v_0}{45 r^4} + \frac{2591}{420} p_{r_{*}}^2 r^3 v_0^{15} + \frac{78314 p_{r_{*}}^2 v_0^3}{189 r^3} \nonumber\\
& - \frac{2}{9} p_{r_{*}}^2 r^2 v_0^{13} - \frac{34}{135} p_{r_{*}}^2 r v_0^{11} + \frac{248 p_{r_{*}}^2 v_0^7}{45 r} - \frac{2104 p_{r_{*}}^2 v_0^9}{35} + \frac{22}{15 r^{10} v_0^9} + \frac{28}{45 r^8 v_0^5} - \frac{1007}{70 r^7 v_0^3} - \frac{2}{27} r^6 v_0^{23} \nonumber\\
& - \frac{562 r^5 v_0^{21}}{315} - \frac{28 v_0}{15 r^5} + \frac{11 r^4 v_0^{19}}{35} + \frac{1209829 v_0^3}{11340 r^4} + \frac{10 r^3 v_0^{17}}{9} + \frac{4027 r^2 v_0^{15}}{1260} - \frac{113 v_0^7}{45 r^2}+ \frac{48 r v_0^{13}}{5} - \frac{12449 v_0^9}{270 r} \nonumber\\ 
& - \frac{31904 v_0^{11}}{567} \ ,
\end{align} 
\begin{equation}
\mF_{\varphi}^{\text{2PN,SS}} = a^2 \left[\frac{116 p_{r_{*}}^4}{15 r^5 v_0^3}+\frac{88 p_{r_{*}}^2}{15 r^6 v_0^3}-\frac{112 p_{r_{*}}^2 v_0}{15 r^4}-\frac{220 p_{r_{*}}^2 v_0^3}{3 r^3}+\frac{2}{5} r^4 v_0^{19}-\frac{409 v_0^3}{15 r^4}+\frac{211 v_0^9}{15 r}+\frac{64 v_0^{11}}{5} \right],
\end{equation}
\begin{align}
\mF_{\varphi}^{\text{2.5PN,Tail}} = &\, \pi \bigg[ \frac{12083  p_{r_{*}}^6}{3780 r^3} - \frac{267293  p_{r_{*}}^4}{2520 r^4} - \frac{10403  p_{r_{*}}^2}{35 r^5} - \frac{64  p_{r_{*}}^2 v_0^4}{3 r^3} - \frac{5  p_{r_{*}}^2 v_0^6}{r^2} + \frac{64}{15}  p_{r_{*}}^2 v_0^{10} - \frac{2039}{70 r^6} + \frac{16}{45} r^4 v_0^{20} \nonumber\\ 
&\quad - \frac{128 v_0^4}{3 r^4} + \frac{22499  v_0^6}{210 r^3} + \frac{64}{15} r^2 v_0^{16} + \frac{64 v_0^{10}}{5 r} - \frac{16621  v_0^{12}}{315} \bigg], 
\end{align} 
\begin{align}
\mF_{\varphi}^{\text{3PN}} = &\, \frac{5 r^8 v_0^{29}}{324} - \frac{1433 r^7 v_0^{27}}{6160} + \frac{173 r^6 v_0^{25}}{1620} + \frac{11}{162} p_{r_{*}}^2 r^6 v_0^{23} - \frac{89 r^5 v_0^{23}}{135} + \frac{1009 p_{r_{*}}^2 r^5 v_0^{21}}{18480} - \frac{65599427 r^4 v_0^{21}}{4989600} \nonumber\\ 
& - \frac{11}{60} p_{r_{*}}^2 r^4 v_0^{19} + \frac{3583 r^3 v_0^{19}}{1260} + \frac{17}{324} p_{r_{*}}^4 r^4 v_0^{17} - \frac{88}{405} p_{r_{*}}^2 r^3 v_0^{17} + \frac{18203 r^2 v_0^{17}}{34020} - \frac{30493 p_{r_{*}}^4 r^3 v_0^{15}}{3080} \nonumber\\ 
& + \frac{209245 p_{r_{*}}^2 r^2 v_0^{15}}{7392} - \frac{306672181 r v_0^{15}}{1188000} + \frac{6848}{175} r v_0^{15} \ln r + \frac{13}{108} p_{r_{*}}^4 r^2 v_0^{13} + \frac{269 p_{r_{*}}^2 r v_0^{13}}{1260} + \frac{109568}{525} v_0^{13} \ln v_0 \nonumber\\ 
& + \frac{54784 \ln 2}{525} v_0^{13} + \frac{2290065817 v_0^{13}}{3118500} - \frac{21307 p_{r_{*}}^2 v_0^{11}}{3402} + \frac{119}{405} p_{r_{*}}^4 r v_0^{11} - \frac{42736 v_0^{11}}{945 r} + \frac{236999 p_{r_{*}}^4 v_0^9}{3168} \nonumber\\
& + \frac{27463 p_{r_{*}}^6 r v_0^9}{2640} - \frac{46224 p_{r_{*}}^2 v_0^9 \ln r}{175 r} + \frac{1712 v_0^9 \ln r}{63 r^2} + \frac{45309763 p_{r_{*}}^2 v_0^9}{23625 r} - \frac{49307783 v_0^9}{79200 r^2} + \frac{133 p_{r_{*}}^6 v_0^7}{324} \nonumber\\
& - \frac{1451 p_{r_{*}}^4 v_0^7}{540 r} + \frac{7427 p_{r_{*}}^2 v_0^7}{540 r^2} - \frac{2269 v_0^7}{180 r^3} + \frac{187 p_{r_{*}}^4 v_0^5}{1620 r^2} + \frac{87736 p_{r_{*}}^2 v_0^5}{1701 r^3} + \frac{873433 v_0^5}{8505 r^4} + \frac{6848 p_{r_{*}}^4 v_0^3 \ln r}{175 r^3} \nonumber\\
& + \frac{232832 p_{r_{*}}^2 v_0^3 \ln r}{1575 r^4} + \frac{1712 v_0^3 \ln r}{45 r^5} - \frac{49153 p_{r_{*}}^8 v_0^3}{18480 r} - \frac{1878557 p_{r_{*}}^6 v_0^3}{184800 r^2} - \frac{3763348417 p_{r_{*}}^4 v_0^3}{8316000 r^3} \nonumber\\ 
& + \frac{2982393643 p_{r_{*}}^2 v_0^3}{8316000 r^4} + \frac{189547637 v_0^3}{1663200 r^5} - \frac{77 p_{r_{*}}^6 v_0}{108 r^3} + \frac{539 p_{r_{*}}^4 v_0}{180 r^4} - \frac{7861 p_{r_{*}}^2 v_0}{540 r^5} - \frac{175 v_0}{54 r^6} + \frac{8959 p_{r_{*}}^8}{5040 r^4 v_0^3} \nonumber\\
& + \frac{1169 p_{r_{*}}^6}{720 r^5 v_0^3} - \frac{480959 p_{r_{*}}^4}{4320 r^6 v_0^3} - \frac{10877563 p_{r_{*}}^2}{45360 r^7 v_0^3} - \frac{279001}{5670 r^8 v_0^3} - \frac{91 p_{r_{*}}^6}{540 r^6 v_0^5} + \frac{7 p_{r_{*}}^4}{108 r^7 v_0^5} + \frac{28 p_{r_{*}}^2}{27 r^8 v_0^5} + \frac{7}{15 r^9 v_0^5} \nonumber\\ 
& - \frac{1189 p_{r_{*}}^8}{1680 r^7 v_0^9} - \frac{9017 p_{r_{*}}^6}{3360 r^8 v_0^9} - \frac{6047 p_{r_{*}}^4}{1120 r^9 v_0^9} - \frac{3041 p_{r_{*}}^2}{420 r^{10} v_0^9} - \frac{509}{168 r^{11} v_0^9} + \frac{7 p_{r_{*}}^6}{324 r^9 v_0^{11}} + \frac{7 p_{r_{*}}^4}{54 r^{10} v_0^{11}} + \frac{7 p_{r_{*}}^2}{27 r^{11} v_0^{11}} \nonumber\\
& + \frac{14}{81 r^{12} v_0^{11}} + \frac{29 p_{r_{*}}^8}{240 r^{10} v_0^{15}} + \frac{49 p_{r_{*}}^6}{60 r^{11} v_0^{15}} + \frac{2 p_{r_{*}}^4}{r^{12} v_0^{15}} + \frac{31 p_{r_{*}}^2}{15 r^{13} v_0^{15}} + \frac{11}{15 r^{14} v_0^{15}} \ , 
\end{align} 
\begin{align}
\mF_{\varphi}^{\text{3PN,Tail}} = & -\frac{7482724 p_{r_{*}}^6 v_0^3}{2625 r^2} + \frac{41796875 p_{r_{*}}^6 v_0^3 \ln 5}{189 r^2} + \frac{34399323 p_{r_{*}}^6 v_0^3 \ln 3}{125 r^2} - \frac{22339490816 p_{r_{*}}^6 v_0^3 \ln 2}{23625 r^2} - \frac{1550216 p_{r_{*}}^4 v_0^3}{525 r^3} \nonumber\\
&  - \frac{1872072 p_{r_{*}}^4 v_0^3 \ln 3}{35 r^3} + \frac{140575744 p_{r_{*}}^4 v_0^3 \ln 2}{1575 r^3} + \frac{1136 \pi ^2 p_{r_{*}}^2 v_0^3}{45 r^4} - \frac{121552 \gamma  p_{r_{*}}^2 v_0^3}{1575 r^4} - \frac{29951111 p_{r_{*}}^2 v_0^3}{55125 r^4} + \frac{16 \pi ^2 v_0^3}{45 r^5} \nonumber\\
& + \frac{60776 p_{r_{*}}^2 v_0^3 \ln r}{525 r^4} - \frac{156006 p_{r_{*}}^2 v_0^3 \ln 3}{175 r^4} + \frac{1111088 p_{r_{*}}^2 v_0^3 \ln 2}{1575 r^4}  - \frac{1712 \gamma  v_0^3}{1575 r^5} + \frac{8511719 v_0^3}{55125 r^5} + \frac{856 v_0^3 \ln r}{525 r^5} \nonumber\\
& - \frac{156006 v_0^3 \ln 3}{175 r^5} + \frac{279056 v_0^3 \ln 2}{315 r^5} - \frac{1552 \pi ^2 v_0^9}{45 r^2}  + \frac{166064 \gamma  v_0^9}{1575 r^2} + \frac{2697337 v_0^9}{55125 r^2} - \frac{83032 v_0^9 \ln r}{525 r^2} + \frac{156006 v_0^9 \ln 3}{175 r^2} \nonumber\\
&  - \frac{152368 v_0^9 \ln 2}{225 r^2} + \frac{512 \pi ^2 v_0^{13}}{15} - \frac{54784 \gamma  v_0^{13}}{525} - \frac{2777632 v_0^{13}}{18375} - \frac{54784}{175} v_0^{13} \ln v_0 - \frac{54784 \ln 2}{175} v_0^{13} \ ,
\end{align} 
\end{subequations}

\subsection{Schott terms} \label{Schott terms}
Here we provide the expressions of the total time derivative of the Schott terms introduced in equations \eqref{Energy balance equations} and \eqref{Angular M. balance equations}. They are expressed as functions of the variables $\{ r, p_{r_{*}}, v_0 \}$.
For the Schott term contribution to the energy balance equation we have
\begin{equation}
  \dot{E}_{\text{Schott}} = \dot{E}_{\text{Sch}}^{\text{0PN}} + \dot{E}_{\text{Sch}}^{\text{1PN}} + \dot{E}_{\text{Sch}}^{\text{1.5PN,Tail}}+ \dot{E}_{\text{Sch}}^{\text{1.5PN,SO}} + \dot{E}_{\text{Sch}}^{\text{2PN}} + \dot{E}_{\text{Sch}}^{\text{2PN,SS}} + \dot{E}_{\text{Sch}}^{\text{2.5PN,Tail}} + \dot{E}_{\text{Sch}}^{\text{3PN}} + \dot{E}_{\text{Sch}}^{\text{3PN,Tail}} \ ,
\end{equation}
where
\begin{subequations}
\begin{equation} \label{Energy_Schott_term_0PN}
  \dot{E}_{\text{Sch}}^{\text{0PN}} = -\frac{152 p_{r_{*}}^4}{15 r^3}-\frac{76 p_{r_{*}}^2}{5 r^4}+\frac{548 p_{r_{*}}^2 v_0^6}{15 r}+\frac{16 v_0^6}{3 r^2}-\frac{16 rv_0^{12}}{3} \ ,
\end{equation}
\begin{equation}
  \dot{E}_{\text{Sch}}^{\text{1PN}} = -\frac{584 p_{r_{*}}^6}{105 r^3}-\frac{20213 p_{r_{*}}^4}{105 r^4}-\frac{1142 p_{r_{*}}^4 v_0^6}{105 r}-\frac{4003 p_{r_{*}}^2}{21 r^5}+\frac{12794 p_{r_{*}}^2 v_0^6}{105 r^2}-\frac{8}{3} p_{r_{*}}^2 r v_0^{12}-\frac{592}{105 r^6}+\frac{278 r^3 v_0^{18}}{105}+\frac{215 v_0^6}{21 r^3}-\frac{761 v_0^{12}}{105} \ , 
\end{equation}
\begin{equation}
  \dot{E}_{\text{Sch}}^{\text{1.5PN,Tail}} = -\frac{50 \pi  p_{r_{*}}^2 v_0^3}{3 r^4}-\frac{10 \pi  v_0^3}{3 r^5}+\frac{10 \pi  v_0^9}{3 r^2} \ , 
\end{equation}
\begin{equation}
  \dot{E}_{\text{Sch}}^{\text{1.5PN,SO}} = a \left[-\frac{48 p_{r_{*}}^4 v_0^3}{5 r^3}-\frac{1708 p_{r_{*}}^2 v_0^3}{15 r^4}-\frac{36 p_{r_{*}}^2 v_0^9}{5 r}-\frac{416 v_0^3}{15 r^5}+\frac{76 v_0^9}{3 r^2}+\frac{12}{5} r v_0^{15} \right] , 
\end{equation}
\begin{align}
  \dot{E}_{\text{Sch}}^{\text{2PN}} = &\, \frac{2593 p_{r_{*}}^8}{315 r^3}+\frac{5633 p_{r_{*}}^6}{45 r^4}+\frac{5368 p_{r_{*}}^6 v_0^6}{315 r}-\frac{224522 p_{r_{*}}^4}{189 r^5}+\frac{70018 p_{r_{*}}^4 v_0^6}{315 r^2}+\frac{199}{21} p_{r_{*}}^4 r v_0^{12}-\frac{1864133 p_{r_{*}}^2}{1890 r^6}+\frac{88}{45} p_{r_{*}}^2 r^3v_0^{18} \nonumber\\ 
  & + \frac{7235 p_{r_{*}}^2 v_0^6}{18 r^3}+\frac{25583 p_{r_{*}}^2 v_0^{12}}{315}-\frac{967}{189 r^7}+\frac{58 r^5 v_0^{24}}{45}-\frac{161113 v_0^6}{1890 r^4}-\frac{388 r^2 v_0^{18}}{35}+\frac{189299 v_0^{12}}{1890 r} \ ,
\end{align}
\begin{equation}
  \dot{E}_{\text{Sch}}^{\text{2PN, SS}} = a^2 \left[-\frac{268 p_{r_{*}}^4}{15 r^5}+\frac{73 p_{r_{*}}^2}{15 r^6}+\frac{1193 p_{r_{*}}^2 v_0^6}{15 r^3}+\frac{211 v_0^6}{15 r^4}-\frac{211 v_0^{12}}{15 r}\right],
\end{equation}
\begin{equation}
  \dot{E}_{\text{Sch}}^{\text{2.5PN, Tail}} = \pi \left[\frac{25 p_{r_{*}}^4}{6 r^7 v_0^3}+\frac{25 p_{r_{*}}^4 v_0^3}{2 r^4}+\frac{55 p_{r_{*}}^2}{6 r^8 v_0^3}+\frac{153397
  p_{r_{*}}^2 v_0^3}{210 r^5}+\frac{35 p_{r_{*}}^2 v_0^9}{3 r^2}+\frac{5}{3 r^9 v_0^3}+\frac{4052 v_0^3}{35
  r^6}-\frac{24487 v_0^9}{210 r^3}-\frac{5 v_0^{15}}{6} \right], 
\end{equation}
\begin{align}
  \dot{E}_{\text{Sch}}^{\text{3PN}} = & -\frac{3277 p_{r_{*}}^{10}}{385 r^3}-\frac{442027 p_{r_{*}}^8}{6930 r^4}-\frac{130111 p_{r_{*}}^8 v_0^6}{6930 r}-\frac{272717041 p_{r_{*}}^6}{1039500 r^5}+\frac{27392 p_{r_{*}}^6 \ln r}{175 r^5}-\frac{1662869 p_{r_{*}}^6 v_0^6}{13860 r^2}-\frac{41786 p_{r_{*}}^6 r v_0^{12}}{3465} \nonumber\\ 
  & -\frac{42507545 p_{r_{*}}^4}{16632 r^6}-\frac{80464 p_{r_{*}}^4 \ln r}{315 r^6}-\frac{146}{55} p_{r_{*}}^4 r^3 v_0^{18}-\frac{1432689791 p_{r_{*}}^4 v_0^6}{2079000 r^3}+\frac{6848 p_{r_{*}}^4 v_0^6 \ln r}{25 r^3}-\frac{109225 p_{r_{*}}^4 v_0^{12}}{1386} \nonumber\\ 
  & -\frac{34926373 p_{r_{*}}^2}{31500 r^7}-\frac{142096 p_{r_{*}}^2 \ln r}{525 r^7}-\frac{3716 p_{r_{*}}^2 r^5 v_0^{24}}{3465}+\frac{3100400579 p_{r_{*}}^2 v_0^6}{1039500 r^4}-\frac{198592 p_{r_{*}}^2 v_0^6 \ln r}{1575 r^4}-\frac{321193p_{r_{*}}^2 r^2 v_0^{18}}{13860} \nonumber\\ 
  & -\frac{127528781 p_{r_{*}}^2 v_0^{12}}{346500 r}+\frac{13696 p_{r_{*}}^2 v_0^{12} \ln r}{175 r}+\frac{880568}{10395 r^8}-\frac{1483 r^7 v_0^{30}}{6930}+\frac{7537561 v_0^6}{38500 r^5}-\frac{142096 v_0^6 \ln r}{1575 r^5}+\frac{1417 r^4 v_0^{24}}{770} \nonumber\\ 
  & -\frac{1133672317 v_0^{12}}{2079000 r^2}+\frac{29104 v_0^{12} \ln r}{225 r^2}+\frac{182383141 r v_0^{18}}{693000}-\frac{6848}{175} r v_0^{18} \ln r \ , 
\end{align}
\begin{align}
  \dot{E}_{\text{Sch}}^{\text{3PN, Tail}} = & -\frac{1552 \pi ^2 p_{r_{*}}^4}{9 r^6}+\frac{166064 \gamma  p_{r_{*}}^4}{315 r^6}+\frac{4441009 p_{r_{*}}^4}{11025 r^6}-\frac{83032 p_{r_{*}}^4 \ln r}{105 r^6}+\frac{156006 p_{r_{*}}^4 \ln 3}{35 r^6}-\frac{152368 p_{r_{*}}^4 \ln 2}{45 r^6} \nonumber\\ 
  & -\frac{4624 \pi ^2 p_{r_{*}}^2}{15 r^7}+\frac{494768 \gamma  p_{r_{*}}^2}{525 r^7}+\frac{27991609 p_{r_{*}}^2}{18375 r^7}-\frac{247384 p_{r_{*}}^2 \ln r}{175 r^7}+\frac{468018 p_{r_{*}}^2 \ln 3}{175 r^7}-\frac{409168 p_{r_{*}}^2 \ln 2}{525 r^7} \nonumber\\ 
  & -\frac{6208 \pi ^2 p_{r_{*}}^2 v_0^6}{45 r^4}+\frac{664256 \gamma  p_{r_{*}}^2 v_0^6}{1575 r^4}+\frac{19507708 p_{r_{*}}^2 v_0^6}{55125 r^4}-\frac{332128 p_{r_{*}}^2 v_0^6 \ln r}{525 r^4}+\frac{624024 p_{r_{*}}^2 v_0^6 \ln 3}{175 r^4} \nonumber\\ 
  & -\frac{609472 p_{r_{*}}^2 v_0^6 \ln 2}{225 r^4}-\frac{512 \pi ^2}{15 r^8}+\frac{54784 \gamma }{525 r^8}+\frac{3736352}{18375 r^8}-\frac{27392 \ln r}{175 r^8}+\frac{109568 \ln 2}{525 r^8}-\frac{16 \pi ^2 v_0^6}{45 r^5} +\frac{1712 \gamma  v_0^6}{1575 r^5} \nonumber\\ 
  &-\frac{8511719 v_0^6}{55125 r^5}-\frac{856 v_0^6 \ln r}{525 r^5}+\frac{156006 v_0^6 \ln 3}{175 r^5}-\frac{279056 v_0^6 \ln 2}{315 r^5}+\frac{1552 \pi ^2 v_0^{12}}{45 r^2}-\frac{166064 \gamma  v_0^{12}}{1575 r^2} -\frac{2697337 v_0^{12}}{55125 r^2} \nonumber\\ 
  &+\frac{83032 v_0^{12} \ln r}{525 r^2}-\frac{156006 v_0^{12} \ln 3}{175 r^2}+\frac{152368 v_0^{12} \ln 2}{225 r^2} \ ,
\end{align}
\end{subequations}
while for the Schott terms related to the angular momentum flux we have
\begin{equation}
  \dot{J}_{\text{Schott}} = \dot{J}_{\text{Sch}}^{\text{0PN}} + \dot{J}_{\text{Sch}}^{\text{1PN}} + \dot{J}_{\text{Sch}}^{\text{1.5PN,Tail}} + \dot{J}_{\text{Sch}}^{\text{1.5PN,SO}} + \dot{J}_{\text{Sch}}^{\text{2PN}} + \dot{J}_{\text{Sch}}^{\text{2PN,SS}} + \dot{J}_{\text{Sch}}^{\text{2.5PN,Tail}} + \dot{J}_{\text{Sch}}^{\text{3PN}} + \dot{J}_{\text{Sch}}^{\text{3PN,Tail}} \ ,
\end{equation}
with
\begin{subequations}
  \begin{equation} \label{Angular_momentum_Schott_term_0PN}
    \dot{J}_{\text{Sch}}^{\text{0PN}} = \frac{256 p_{r_{*}}^2 v_0^3}{15 r}+\frac{128 v_0^3}{15 r^2}-\frac{128 r v_0^9}{15} \ ,
  \end{equation}
  \begin{equation}
    \dot{J}_{\text{Sch}}^{\text{1PN}} = -\frac{64 p_{r_{*}}^4}{15 r^4 v_0^3}-\frac{64 p_{r_{*}}^4 v_0^3}{15 r}-\frac{32 p_{r_{*}}^2}{3 r^5 v_0^3}-\frac{6227 p_{r_{*}}^2 v_0^3}{105 r^2}-\frac{32}{5} p_{r_{*}}^2 r v_0^9-\frac{64}{15 r^6 v_0^3}-\frac{32}{15} r^3 v_0^{15}-\frac{219 v_0^3}{35 r^3}+\frac{443 v_0^9}{35} \ ,
  \end{equation}
  \begin{equation}
    \dot{J}_{\text{Sch}}^{\text{1.5PN,Tail}} = -\frac{10 \pi  p_{r_{*}}^2}{r^4}-\frac{10 \pi }{3 r^5}+\frac{10 \pi  v_0^6}{3 r^2} \ , 
  \end{equation}
  \begin{equation}
    \dot{J}_{\text{Sch}}^{\text{1.5PN,SO}} = a \left[ -\frac{16 p_{r_{*}}^4}{r^3}-\frac{448 p_{r_{*}}^2}{5 r^4}-\frac{24 p_{r_{*}}^2 v_0^6}{5 r}-\frac{152}{5 r^5}+\frac{116 v_0^6}{5 r^2}+\frac{36}{5} r v_0^{12} \right],
\end{equation}
  \begin{align}
    \dot{J}_{\text{Sch}}^{\text{2PN}} = & -\frac{8 p_{r_{*}}^6}{15 r^7 v_0^9}+\frac{64 p_{r_{*}}^6}{15 r^4 v_0^3}+\frac{8 p_{r_{*}}^6 v_0^3}{3 r}-\frac{12 p_{r_{*}}^4}{5 r^8 v_0^9}+\frac{2113 p_{r_{*}}^4}{140 r^5 v_0^3}+\frac{5443 p_{r_{*}}^4 v_0^3}{420 r^2}+4 p_{r_{*}}^4 r v_0^9-\frac{16 p_{r_{*}}^2}{5 r^9 v_0^9}+\frac{1681 p_{r_{*}}^2}{60 r^6 v_0^3}+\frac{16}{15} p_{r_{*}}^2 r^3 v_0^{15} \\ 
    & -\frac{1487299 p_{r_{*}}^2 v_0^3}{3780 r^3}-\frac{71 p_{r_{*}}^2 v_0^9}{105}-\frac{16}{15 r^{10} v_0^9} +\frac{881}{210 r^7 v_0^3}-\frac{4}{15} r^5 v_0^{21}-\frac{84881 v_0^3}{756 r^4}-\frac{2143 r^2 v_0^{15}}{420}+\frac{216437 v_0^9}{1890 r} \ , 
  \end{align}
  \begin{equation} \label{Angular_momentum_Schott_term_SSPN}
      \dot{J}_{\text{Sch}}^{\text{2PN,SS}} = a^2 \left[-\frac{128 p_{r_{*}}^4}{15 r^5 v_0^3}-\frac{64 p_{r_{*}}^2}{15 r^6 v_0^3}+\frac{1292 p_{r_{*}}^2 v_0^3}{15 r^3}+\frac{307 v_0^3}{15 r^4}-\frac{307 v_0^9}{15 r} \right],
  \end{equation}
  \begin{equation}
    \dot{J}_{\text{Sch}}^{\text{2.5PN,Tail}} = \pi \left[\frac{5  p_{r_{*}}^4}{r^4}+\frac{46048 p_{r_{*}}^2}{105 r^5}+\frac{5 p_{r_{*}}^2 v_0^6}{r^2}+\frac{22499}{210 r^6}-\frac{22499 v_0^6}{210 r^3} \right], 
  \end{equation}
  \begin{align}
    \dot{J}_{\text{Sch}}^{\text{3PN}} = & -\frac{2 p_{r_{*}}^8}{15 r^{10} v_0^{15}}+\frac{14 p_{r_{*}}^8}{15 r^7 v_0^9}-\frac{62 p_{r_{*}}^8}{15 r^4 v_0^3}-\frac{2 p_{r_{*}}^8 v_0^3}{r}-\frac{13 p_{r_{*}}^6}{15 r^{11} v_0^{15}}+\frac{11603 p_{r_{*}}^6}{3360 r^8 v_0^9}-\frac{2923 p_{r_{*}}^6}{420 r^5 v_0^3}-\frac{27551 p_{r_{*}}^6 v_0^3}{3360 r^2}-\frac{49}{15} p_{r_{*}}^6 r v_0^9 \nonumber\\ 
    & -\frac{2 p_{r_{*}}^4}{r^{12} v_0^{15}}+\frac{20189 p_{r_{*}}^4}{3360 r^9 v_0^9}+\frac{3187889 p_{r_{*}}^4}{30240 r^6 v_0^3}-p_{r_{*}}^4 r^3 v_0^{15}+\frac{481217 p_{r_{*}}^4 v_0^3}{4320 r^3}+\frac{399 p_{r_{*}}^4 v_0^9}{160}-\frac{28 p_{r_{*}}^2}{15 r^{13} v_0^{15}}+\frac{1273 p_{r_{*}}^2}{210 r^{10} v_0^9} \nonumber\\ 
    & +\frac{56683 p_{r_{*}}^2}{240 r^7 v_0^3}+\frac{1}{3} p_{r_{*}}^2 r^5 v_0^{21}+\frac{675397039 p_{r_{*}}^2 v_0^3}{1663200 r^4}-\frac{70192 p_{r_{*}}^2 v_0^3 \ln r}{315 r^4}+\frac{2713 p_{r_{*}}^2 r^2 v_0^{15}}{1120}+\frac{2011469 p_{r_{*}}^2 v_0^9}{15120 r}-\frac{8}{15 r^{14} v_0^{15}} \nonumber\\ 
    & +\frac{219}{280 r^{11} v_0^9}+\frac{102103}{1890 r^8 v_0^3}+\frac{r^7 v_0^{27}}{15}-\frac{121954519 v_0^3}{1188000 r^5}-\frac{70192 v_0^3 \ln r}{1575 r^5}-\frac{449 r^4 v_0^{21}}{224}+\frac{302957833 v_0^9}{8316000 r^2}+\frac{70192 v_0^9 \ln r}{1575 r^2} \nonumber\\ 
    & +\frac{420059 r v_0^{15}}{30240} \ ,
  \end{align}
  \begin{align}
    \dot{J}_{\text{Sch}}^{\text{3PN, Tail}} = & -\frac{1552 \pi ^2 p_{r_{*}}^2 v_0^3}{9 r^4}+\frac{166064 \gamma  p_{r_{*}}^2 v_0^3}{315 r^4}+\frac{4441009 p_{r_{*}}^2 v_0^3}{11025 r^4}-\frac{83032 p_{r_{*}}^2 v_0^3 \ln r}{105 r^4}+\frac{156006 p_{r_{*}}^2 v_0^3 \ln 3}{35 r^4}-\frac{1552 \pi ^2 v_0^3}{45 r^5} \nonumber\\ 
    & -\frac{152368 p_{r_{*}}^2 v_0^3 \ln 2}{45 r^4}+\frac{166064 \gamma  v_0^3}{1575 r^5}+\frac{2697337 v_0^3}{55125 r^5}-\frac{83032 v_0^3 \ln r}{525 r^5}+\frac{156006 v_0^3 \ln 3}{175 r^5}  -\frac{152368 v_0^3 \ln 2}{225 r^5} \nonumber\\ 
    &+\frac{1552 \pi ^2 v_0^9}{45 r^2}-\frac{166064 \gamma  v_0^9}{1575 r^2}-\frac{2697337 v_0^9}{55125 r^2}+\frac{83032 v_0^9 \ln r}{525 r^2}-\frac{156006 v_0^9 \ln 3}{175 r^2} +\frac{152368 v_0^9 \ln 2}{225 r^2}.    
  \end{align}
\end{subequations}
We remark that these total time derivatives of the Schott terms have to be computed on the geodesics and plugged in the balance equations.

\begin{figure*} 
  \includegraphics[width=1.\linewidth]{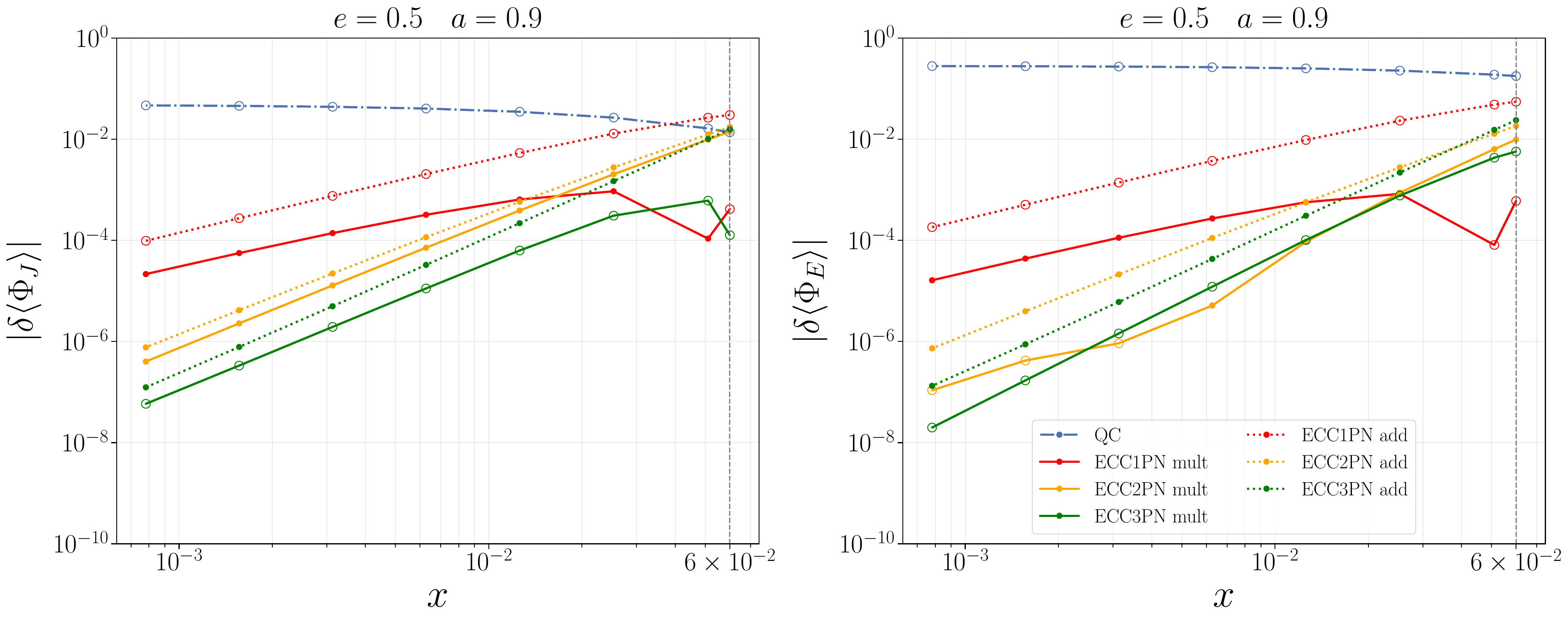}
  \caption{Absolute value of the fractional differences defined in Eq.~\eqref{Eq: fractional difference definition} of the averaged fluxes computed on orbits in a weak-field regime. 
  The x-axis corresponds to the gauge-invariant variable $x = \langle \dot{\varphi} \rangle^{2/3}$. The orbits are characterized by the parameters $e = 0.5$, $a = 0.9$ and $p = \{ 960, 480, 240, 120, 60, 30, 15, 12.797 \}$. 
  Dotted lines correspond to the additive implementation (add) while solid lines correspond to the multiplicative (mult) implementation. Empty circle dots represent negative values of the fractional differences, highlighting zero crossing in the log-log plots.
  We highlight with a dashed vertical gray line the orbit with $x = 0.06$, which is also shown in the contour plot of Fig.~\ref{spin and ecc}.}
  \label{Fig: averaged-fluxes-spinning WF}
\end{figure*}
\end{widetext}

\subsection{PN scaling assessment for spinning orbits} \label{PN scaling assessment for spinning orbits}

In Fig.~\ref{Fig: averaged-fluxes-spinning WF}, we show a similar plot to the one in Fig.~\ref{Fig: averaged-fluxes WF} for eccentric orbits with $e = 0.5$ around a Kerr BH with spin $a = 0.9$.
We test the weak-field regime scaling of the different PN order of the eccentric corrections to the RR force in Eq.~\eqref{ecc impl} by studying the fractional differences of the orbit-averaged 
analytical and numerical fluxes.
As in the non-spinning case, we recover the correct scaling of the different PN truncations, which is expected to be $x^{-(n+1/2)}$, with $n$ being the PN order at which we perform the truncation. However, in the spinning case, our analysis has an important caveat: 
the analytical fluxes we consider contain SO and SS spin corrections at 2PN. This means that the ECC3PN fluxes do not contain the spin contributions at 2.5PN and 3PN. 
As a consequence the ECC3PN lines in Fig.~\ref{Fig: averaged-fluxes-spinning WF} exhibit the same scaling (slope) as the ECC2PN, 
but they still improve the ECC2PN lines exhibiting a shift down. In the plot we highlight with a vertical gray curve the orbit with $x = 0.06$ which is present in the contour plot of Fig.~\ref{spin and ecc}. 
This plot extends to regimes $x \le 0.06$ what we found in Fig.~\ref{spin and ecc}: the multiplicative ECC3PN corrections in Eq.~\eqref{multiplicative impl} perform better than the additive ECC3PN in Eq.~\eqref{multiplicative impl}, especially for eccentric orbits with prograde spin. 

\section{Instantaneous fluxes from frequency domain Teukolsky solutions}\label{App:FDInstFlux}
\newcommand{\SWSH}[3][]{\,{_{#2}S_{#3}^{\mathrm{#1}}}}	%Spin-weighted spheroidal harmonic
\newcommand{\Y}[3][]{\,{_{#2}Y_{#3}^{\mathrm{#1}}}}		%Spin-weighted spherical harmonic
\newcommand{\Sb}[4]{({_{#1}b_{#2}})^{#3}_{#4}}			%spheroidal-to-spherical projector

The frequency domain solutions to the Teukolsky equation at future null infinity for a particle traveling on an eccentric geodesic take the form
\begin{equation}
\psi_4 = \sum_{l m n} Z_{lmn} {_{-2}S_{l m \omega_{mn}}}(\cos\theta) e^{i m \phi}e^{-i\omega_{mn} t},
\end{equation}
where the ${_{-2}S_{l m \omega_{mn}}}$ spin-weighted spheroidal harmonics of weight -2, and the mode frequencies are $\omega_{mn} = m\Omega_\phi+n\Omega_r$ with $\Omega_\phi$ and $\Omega_r$ the azimuthal and radial frequencies. To build the instantaneous fluxes using Eqs.~\eqref{eq:inst_fluxes} we first project the spheroidal harmonics ${_{-2}S_{l m \omega_{mn}}}$ onto spherical harmonics ${_{-2}Y_{l m}}$ using,
\begin{equation}
{_{-2}S_{l m \omega_{mn}}}(\cos\theta) e^{i m \phi} = \sum_{\ell} \Sb{-2}{mn}{\ell}{l} \Y{-2}{\ell m}(\theta,\phi),
\end{equation}
where the coefficients $\Sb{-2}{mn}{\ell}{l}$ are obtained using the algorithm of Ref.~\cite{Hughes:1999bq}. This allows us to define the spin-weighted spherical harmonic coefficients of $\psi_4$,
\begin{equation}
\mathcal{Z}_{\ell mn} = \sum_{l} Z_{lmn}\Sb{-2}{mn}{\ell}{l}.
\end{equation}
Combining the above expressions with Eqs.~\eqref{eq:inst_fluxes}, and applying the orthogonality relations for products of spin-weighted harmonics when integrated over the sphere, we obtain the following closed-form expressions for the instantaneous flux,
\begin{subequations}
	\begin{align}
		\Phi_{E} &= \frac{1}{4\pi}\sum_{\ell m nn'}\frac{ \mathcal{Z}_{\ell mn}\mathcal{Z}_{\ell mn'}^{*} }{\omega_{mn}\omega_{mn'}}e^{-i(n-n')\Omega_r t}\text{, and}\\
		\Phi_{J} &= \frac{1}{4\pi}\sum_{\ell m nn'}m\frac{ \mathcal{Z}_{\ell mn}\mathcal{Z}_{\ell mn'}^{*} }{\omega_{mn}\omega_{mn'}^2}e^{-i(n-n')\Omega_r t}.
	\end{align}
\end{subequations}
\bibliography{references}
\end{document}